\newcommand{\ee}{e^{+}e^{-}}
\newcommand{\mm}{\mu^{+}\mu^{-}}

\newcommand{\mz}{M_{Z}}

\newcommand{\mrec}{M_{\texttt{rec}}}
\newcommand{\mmu}{M_{\mu^{+}\mu^{-}}}
\newcommand{\ms}{M_{S^0}}

\documentclass[11pt,a4paper]{scrartcl}
\usepackage{ILD}

\usepackage{lineno}
\usepackage{tabularx}
\usepackage{subcaption}
\usepackage{amsmath}
\usepackage{amssymb}
\usepackage{graphicx}
\usepackage{xcolor}
\usepackage{multirow}

\title{ILD Benchmark: Search for Extra Scalars Produced in Association with a $Z$ boson at $\sqrt{s}=500$\,GeV}

\date{\today}

\addauthor{Yan Wang}{\institute{1}\institute{2}}
\addauthor{Mikael Berggren}{\institute{1}}
\addauthor{Jenny List}{\institute{1}}

\addinstitute{1}{Deutsches Elektronen-Synchrotron DESY}
\addinstitute{2}{College of Physics and Electronic Information, Inner Mongolia Normal University, Hohhot 010022, China}

\abstract{
We study the prospects for discovering an extra scalar boson $S^0$ at the International Linear Collider (ILC) based on a full simulation of the International Large Detector (ILD). In order to provide results in an as model-independent way as possible, the analysis uses the recoil of the scalar against a $Z$ boson decaying into a pair of muons, $e^+e^- \to \mu\mu S^{0}$. This process serves as a physics benchmark for the ILD detector performance at $\sqrt{s}=500$\,GeV, specifically for the muon ID and momentum resolution, as well as for the identification of initial state radiation photons and their energy measurement. As final results, the sensitivities for discovering the extra scalars at 2 $\sigma$ level are evaluated in terms of a scale factor $\sin^2{\theta}$ with respect to the Standard Model value of the cross section for the Higgs--strahlung process. 
Two detector models, IDR-L and IDR-S, are considered in the
analysis, which differ in radius of the tracking volume, aspect ratio and strength of the magnetic field. While the two detector models show a visible difference in the precision of the reconstructed invariant di-muon mass, no difference is found at the level of the final results.}

\titlecomment{This work was carried out in the framework of the ILD concept group}

\begin{document}

% generates the title page
\titlepage

\section{Introduction}
The motivation of this paper is to study the impact of the detector performance on
the sensitivity of a future Linear Collider to additional Higgs bosons or other new scalar particles, denoted here generically with $S^0$. Specifically, the detector performance in terms of the muon identification and momentum resolution, as well as for the identification of initial state radiation (ISR) photons and their energy measurement will be studied. Since the $125$-GeV Higgs boson is rather SM-like, the coupling of any additional scalar to the $Z$ boson, $ZZS^0$, is expected to be small~\cite{ATLAS, aggleton}. Extra scalars have been searched for previously at LEP~\cite{Barate:2003sz, Schael:2006cr} and LHC~\cite{Aad:2015wra, Khachatryan:2015lba, Khachatryan:2016are, Aaboud:2018eoy, Sirunyan:2019xls, Sirunyan:2019wrn}. However, in most cases, the searches depend on model-specific assumptions on the properties of $S^0$. A notable exception is the recoil search  by the OPAL experiment, which is the basis of the decay-mode independent measurement of the total Higgsstrahlungs cross section~\cite{Abbiendi}. However, the OPAL results were limited by the LEP center-of-mass energy and the relatively low luminosity. 

The International Linear Collider (ILC) is a proposed electron-positron linear collider, whose luminosity will be over a thousand
times higher than that of LEP, which makes the recoil mass technique a more powerful tool in the search for extra scalars~\cite{Asner}. The ILC can reach higher center-of-mass energies and higher luminosities, which will cover wider search regions for the extra scalars. A generator-level extrapolation of the LEP results to ILC at $\sqrt{s}=250$\,GeV has been presented in~\cite{theo}. Preliminary results for ILC operation at 250 GeV and 500 GeV based on full simulation of the ILD dector concept have been reported at LCWS2017~\cite{Wang:2018fcw}, LCWS2018~\cite{Wang:2019mzd} and ICHEP2018~\cite{Wang:2018ichep}. The ILD results for $\sqrt{s}$=500 GeV are superseded by this document.

The $e^{+} e^{-}\to S^{0}Z \to \mu^{+}\mu^{-}S^{0}$ channel is chosen as a benchmark signal in this analysis, since $Z\to \mu^{+}\mu^{-}$ is the cleanest decay mode of the $Z$ boson. Other visible decay modes, like $Z\to e^{+}e^{-}$ or $Z\to q\bar{q}$ can in principle also be used in this search~\cite{Thomson:2015jda,Yan:2016xyx}. The final results will be expressed in terms of $\sin^2{}\theta$, where $\sin{\theta}$ is the ratio of the $ZZS^0$ coupling over the $ZZH$ coupling for a SM Higgs boson of the same mass as the $S^0$. 

In this analysis, important detector performance aspects are the identification and momentum measurement of the two muons, as well as the identification and energy measurement of ISR photons within the detector acceptance. ISR occurs frequently for the lower scalar masses, where $\ms + \mz \ll \sqrt{s}$. If the ISR photon is detected, then the event kinematics can be corrected accordingly, which improves the separation of signal and background significantly.

\section{Event Generation and Detector Simulation}
The Monte-Carlo event samples used in this study have been produced in context of the ILD Interim Design Report (IDR)~\cite{ILD:2020qve}, based on the generator-level events created for the ILC TDR~\cite{Behnke:2013lya} by the LCC Generator Group:  
All the event samples are generated with 100\% left-handed and right-handed beam polarizations, using the $\textsc{Whizard}$~1.95~\cite{Kilian:2007gr} for the hard event, which then passed to $\textsc{Pythia}$~6~\cite{Sjostrand:2006za} for hadronization. 
Then the samples are reweighted with beam polarizations of $\pm 80\%$ for the electron beam and $\pm 30\%$ for the positron beam. The total integrated luminosity of $4000$\,fb$^{-1}$ at $\sqrt{s}=500$\,GeV is shared between the four polarisation sign configurations with fractions $f(-+, +-, ++, --) = (40\%, 40\%, 10\%, 10\%)$.

The signal events have been generated specifically for this analysis, using the same setup as the SM Higgs samples from the DBD production~\cite{Behnke:2013lya}, changing only the mass of the Higgs boson, while keeping the same branching ratios as predicted by the SM for the $125$\,GeV Higgs boson. However, no use is made of this assumption in the analysis. The cross section for $ZS^0$ production assumed in event generation equals the cross section for $ZH$ production predicted in the SM for $M_H = \ms$, i.e.\ $\sin{\theta}=1$.
A total of $48$ different values for $\ms$ are considered in the range of $10 \leq \ms \leq 410$\,GeV. The benchmark points are generally selected every $10$\,GeV; near $\ms = \mz$ and $\ms = M_H$ as well as near the kinematic limit, the interval becomes $5$\,GeV; in mass ranges far from any SM resonance, where recoil mass distribution of the background is rather flat, the interval becomes $20$\,GeV. The full list of benchmark masses is shown in Table~\ref{table_bps}.

\begin{table}[htb]
 \begin{center}
 \begin{small}
 \begin{tabular}{|c| c| c| c| c| c| c| c| c| c| }
 \hline
      10 & 20 & 30 & 40 & 50 & 60 & 70 & 80 & 85 & 90  \\
 \hline
       95 & 100 & 105 & 110 & 115 & 120 & 130 & 135 & 140 & 160  \\
 \hline
       180 & 200 & 220 & 240 & 260 & 280 & 290 & 300 & 310 & 320  \\
 \hline
       325 & 330 & 335 & 340 & 345 & 350 & 355 & 360 & 365 & 370  \\
 \hline
       375 & 380 & 385 & 390 & 395 & 400 & 405 & 408 & - &  - \\
 \hline
 \end{tabular}
 \end{small}
 \end{center}
  \caption{Mass values in GeV of the signal samples used in the analysis.}\label{table_bps}
 \end{table}

All event samples have been simulated, reconstructed and analysed with $\textsc{ILCSoft-v02-00-02}$~\cite{ilcsoft:2019}.  
The events are simulated with the two detector configurations studied in the IDR, IDR-L (ILD{\_}l5{\_}o1{\_}v02) and IDR-S (ILD{\_}s5{\_}o1{\_}v02). IDR-L is very similar to ILD as described in the ILC TDR~\cite{Behnke:2013lya}, while IDR-S features a smaller tracker radius and an increased magnetic field strength. The detailed description of both models can be found in the IDR~\cite{ILD:2020qve}. The $\gamma\gamma\to$ low $p_T$ hadrons and $e^+e^-$ seeable pair background are overlaid to all MC samples before the reconstruction. Then all the events are reconstructed using the PandoraPFA~\cite{Thomson:2009rp} algorithm in the \textsc{Marlin} framework~\cite{Wendt:2007iw} to reconstruct individual final state particles, so-called Particle Flow Objects (PFOs).

\section{Event selection}
The analysis proceeds as follows: First, two oppositely charged isolated muon candidates are identified, which have to fulfill a loose preselection described in Sec.~\ref{subsec:presel}. At this stage, we present comparisons of the di-muon invariant mass measurement for IDR-L and IDR-S in Sec.~\ref{subsec:perform}. Next, ISR (c.f.\ Sec.~\ref{subsec:ISR}) and bremsstrahlung/FSR (c.f.\ Sec.~\ref{subsec:FSR}) photon candidates are identified. The FSR photons are re-combined with the measured muon momenta, and the event kinematics are corrected for identified ISR candidates. Based on these corrected kinematics, two boosted decision trees are trained against the two most important background classes as described in Sec.~\ref{subsec:MVA}. Finally, we discuss the main selection, the cut flow and the composition of the remaining background in Sec.~\ref{subsec:select}. 

\subsection{Muon Identification}
\label{subsec:presel}
Isolated muons are selected by the \textsc{Marlin} processor \textsc{IsolatedLeptonTagging}~\cite{junping:ilt}. In this processor, 
muons are identified mainly by their MIP signal in the calorimeters, requiring the energy deposited in the calorimeters to be less than half of the track momentum.  The isolation is performed by an MVA based on a double-cone method\footnote{The double-cone method exploits various observables based on the energy deposits located within two cones of different sizes around the muon candidate.}. A muon candidate is accepted if the MVA output is larger than $0.8$ and if its momentum is larger than $10$\,GeV. 
Each event is required to have at least one $\mu^+$ and one $\mu^-$ fulfilling these criteria.

Then, if there are more than one $\mu^+$ and one $\mu^-$ candidate, a pair of oppositely charged muons is selected by minimizing the following $\chi^{2}$ function:
\begin{equation}
\chi^{2} (\mmu, \mrec) =
\frac{(\mmu-\mz)^2}{\sigma^{2}_{\mmu}} + 
\frac{(\mrec-\ms)^2}{\sigma^{2}_{\mrec}},
\end{equation} 
where $\mmu$ is the invariant mass of the muon pair, and ${\mrec}$ is the recoil mass, which is defined as:
\begin{equation}
\mrec^2 = (\sqrt{s}-E_{\mu^+\mu^-})^2-|\vec{p}_{\mu^+\mu^-}|^2.
\end{equation} 
$\sigma_{\mmu}$  and $\sigma_{\mrec}$ are the detector resolutions of $\mmu$ and ${\mrec}$, which are calculated event-by-event from the covariance matrix of the track fit.  A very loose preselection is used at this stage for choosing the muon pair: $\mmu\in[\mz-40,\mz+40]$ GeV, $\mrec\in [0,500]$ GeV.

\subsection{Di-muon Invariant Mass Reconstruction Performance}
\label{subsec:perform}
%In the first step after the preselection, kinematic variables that depend only on the oppositely charged muon pair (thus the Z boson candidate) are exploited: the invariant mass and transverse momentum of the muon pair.
Since the di-muon invariant mass and its uncertainty as reconstructed from the measured tracks are important observables in this analysis, we compare the performance of IDR-L and IDR-S in terms of these quantities.

\begin{figure}[htbp]
\begin{center}
 \begin{subfigure}{0.475\textwidth}
    \includegraphics[width=\textwidth]{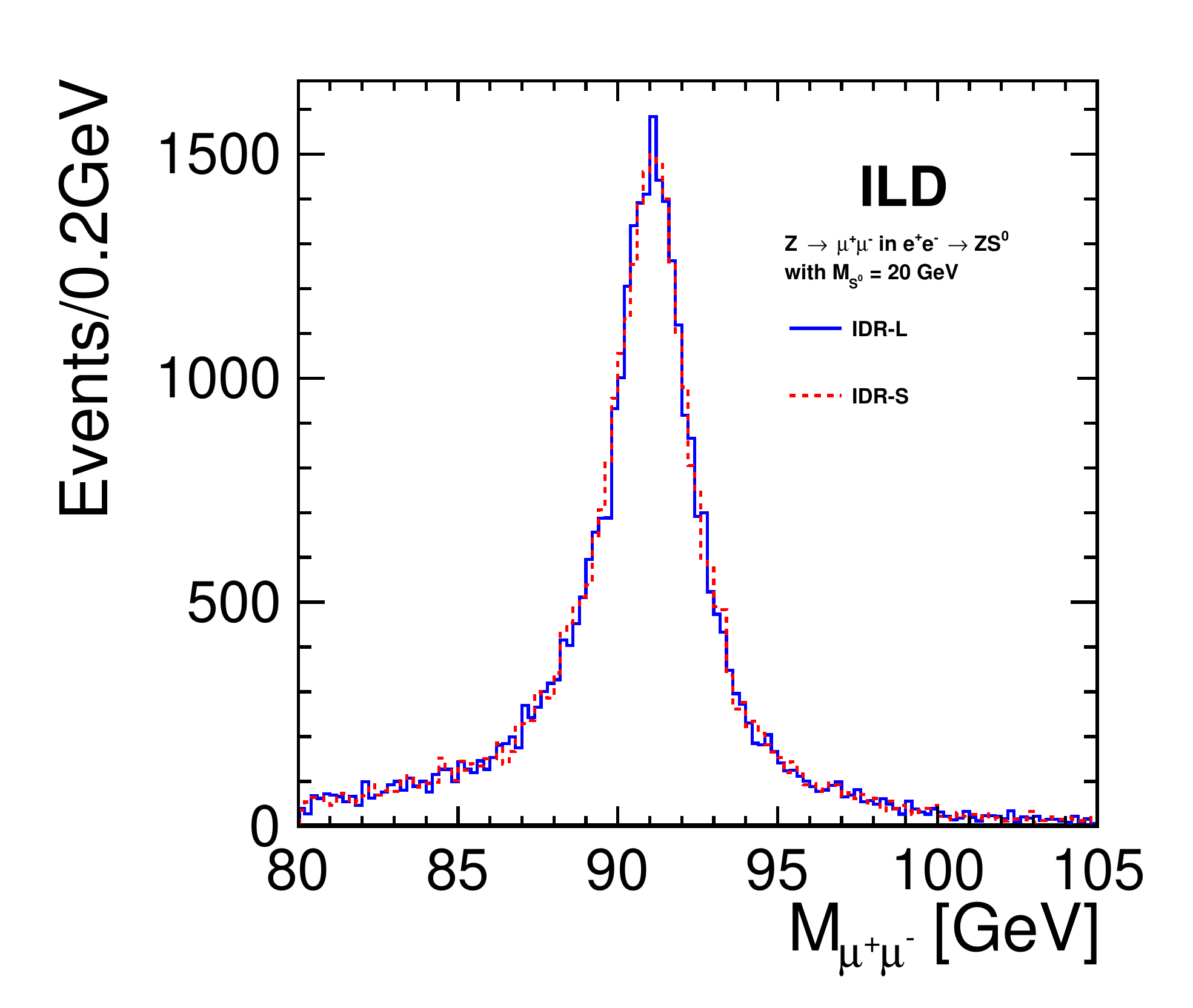} 
    \subcaption{\label{fig_inm:S20}}
 \end{subfigure} 
\hspace{0.03\textwidth}
 \begin{subfigure}{0.475\textwidth}
   \includegraphics[width=\textwidth]{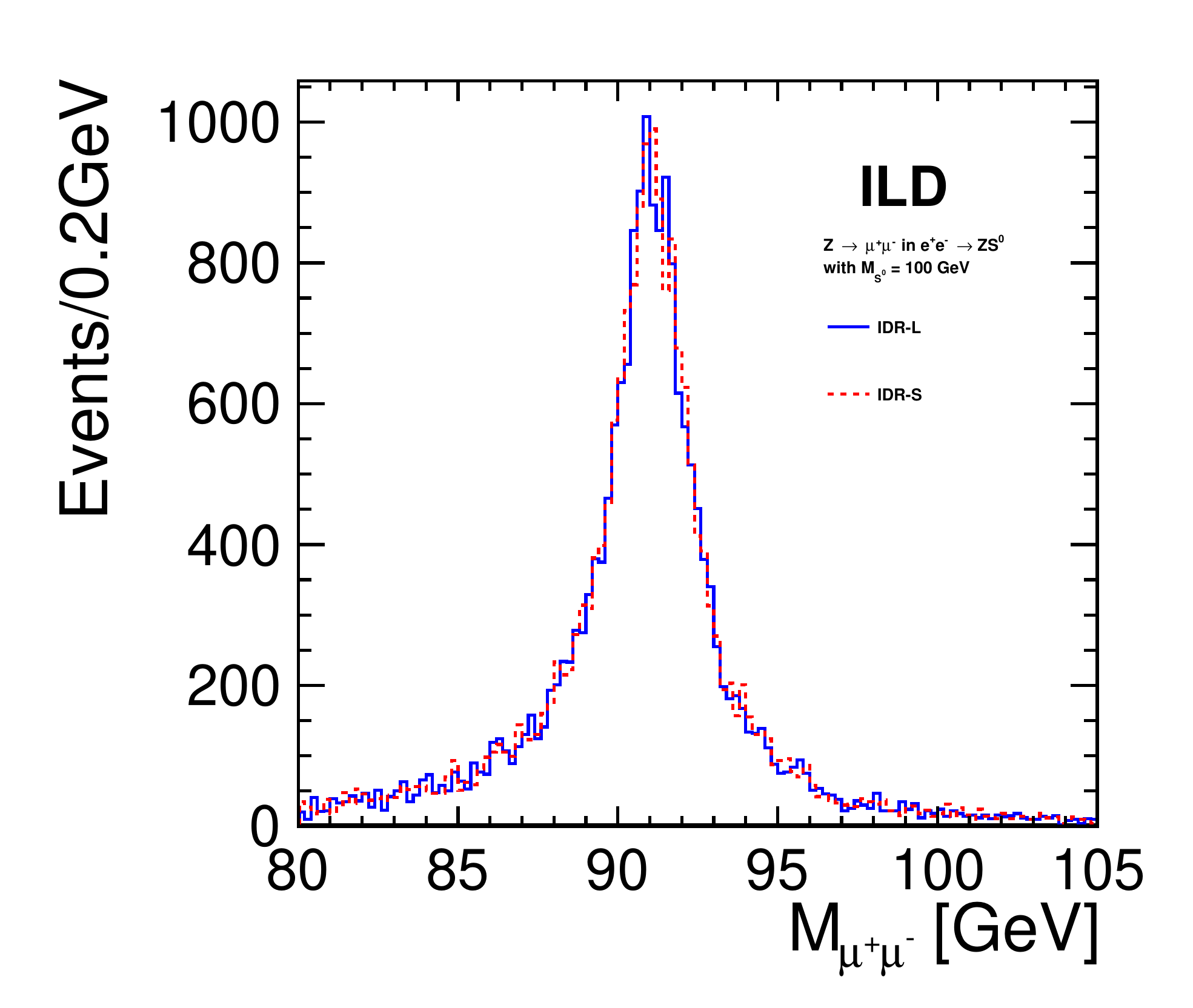} 
   \subcaption{\label{fig_inm:S100}}	
 \end{subfigure} 
 \begin{subfigure}{0.475\textwidth}
   \includegraphics[width=\textwidth]{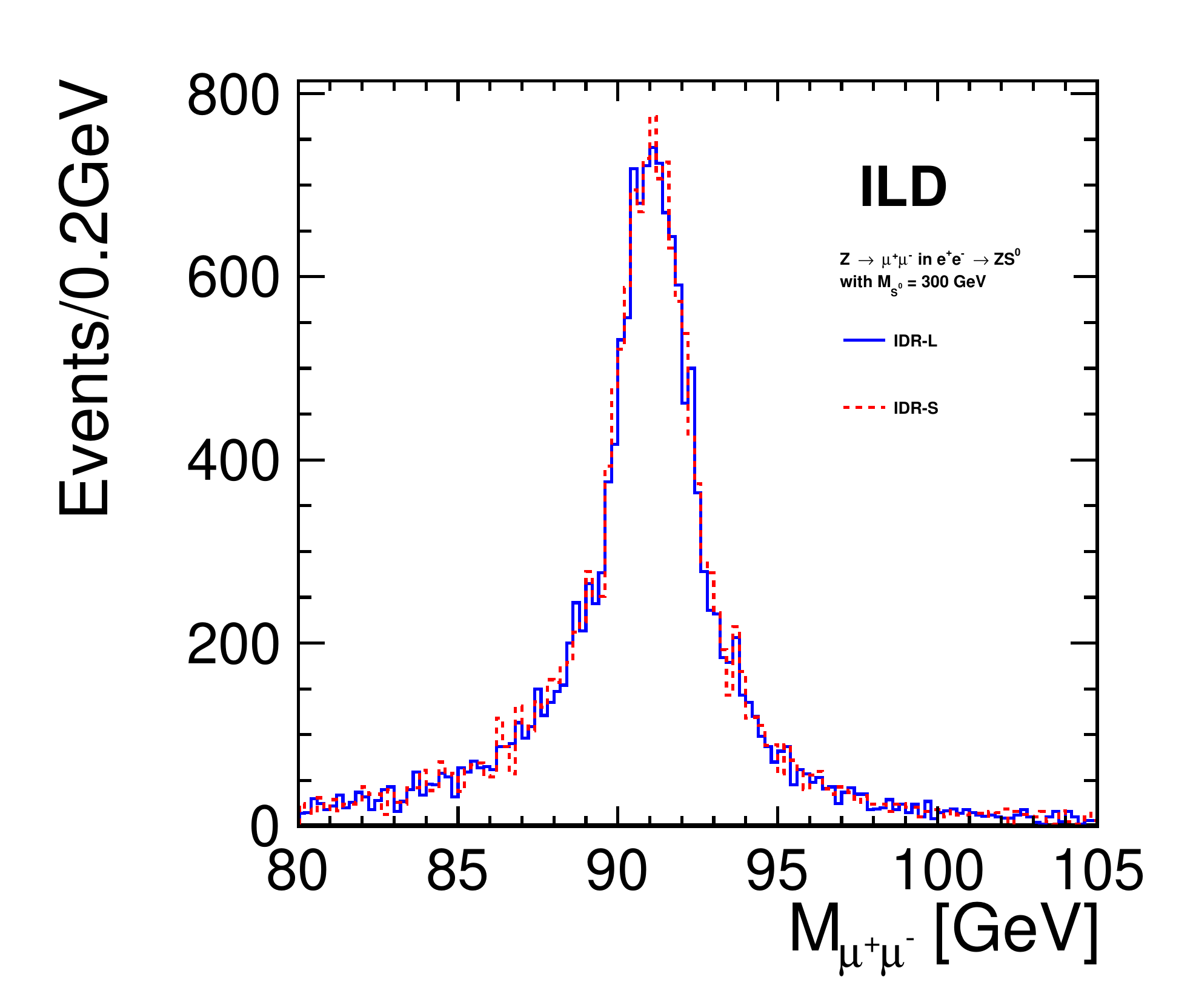} 
    \subcaption{\label{fig_inm:300}}
 \end{subfigure} 
\hspace{0.03\textwidth}
\begin{subfigure}{0.475\textwidth}
   \includegraphics[width=\textwidth]{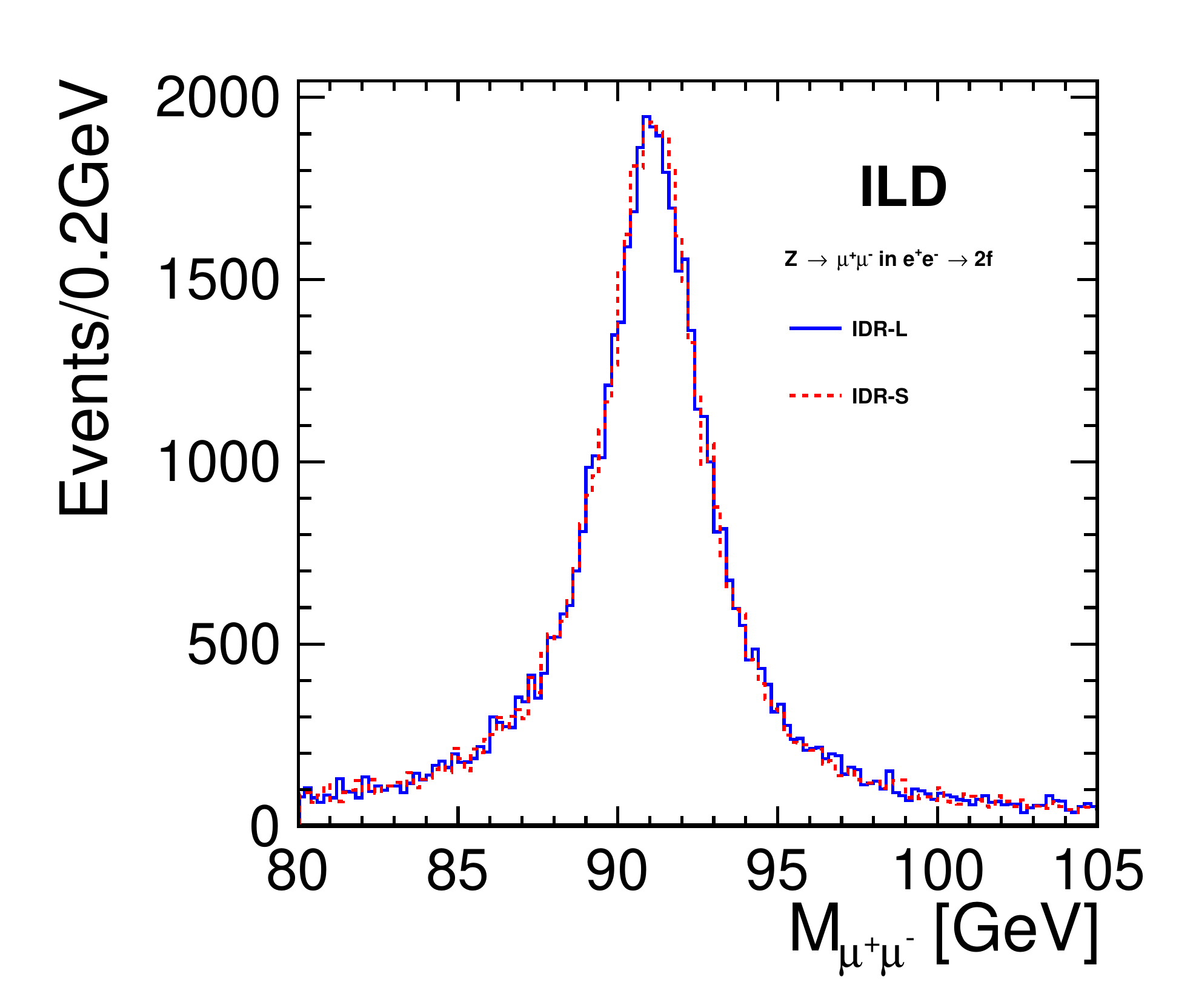} 
   \subcaption{\label{fig_inm:2f}}	
 \end{subfigure}
\end{center}
 \caption{Distributions of the muon pair invariant mass $M_{\mu^{+}\mu{-}}$ in IDR-L and IDR-S:
(a) $ZS^0$ signal with $\ms = 20$\,GeV
(b) $ZS^0$ signal with $\ms = 100$\,GeV
(c) $ZS^0$ signal with $\ms = 300$\,GeV
(d) $2f$ background. }
\label{fig_inm}
\end{figure}

Figure~\ref{fig_inm} shows the di-muon invariant mass distribution for signal events with different scalar masses ($\ms=20,~100,~300$\,GeV) as well as for the two fermion background as reconstructed in IDR-L and IDR-S. In all cases, the distributions show a clear $Z$ peak, with a width not far from the natural width, and with a small asymmetric tail towards lower masses due to the not yet recovered FSR and bremsstrahlung. At this level, no significant difference between the two detector models can be seen.

%There are clear differences between IDR-L and IDR-S when $\ms$ is large. The difference mainly comes from the detector effect, beacause there is the angular dependency in the transverse momentum resolution. In general, the IDR-L has better resolution in the barrel region, while IDR-S has better resolution in the forward region \cite{ILD:2020qve}. 

\begin{figure}[htbp]
\begin{center}
\begin{subfigure}{0.475\hsize} 
\includegraphics[width=\textwidth]{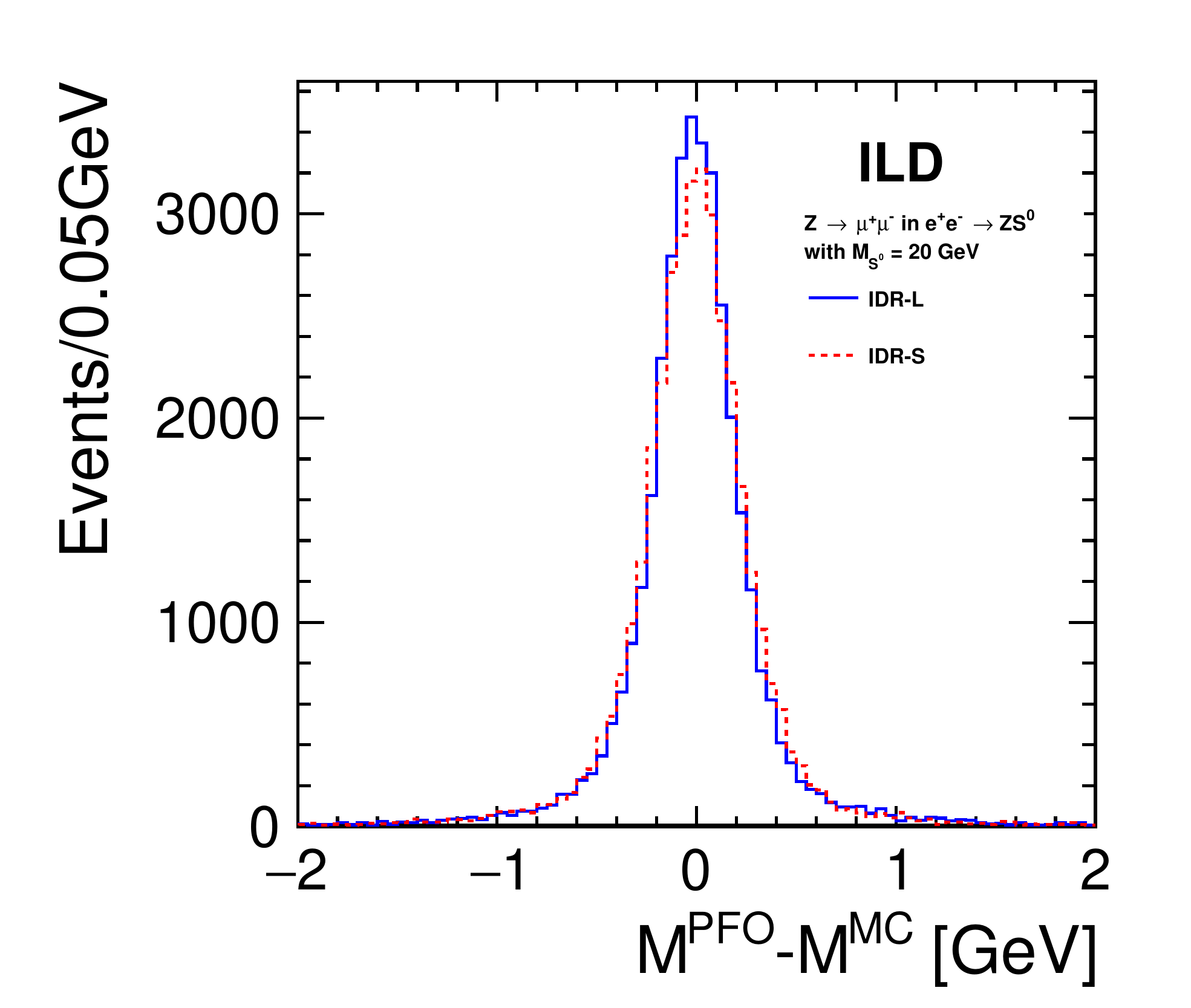}
 \caption{ \label{fig:extraH:Mdiff:mh20}}
 \end{subfigure}
\hspace{0.03\textwidth}
\begin{subfigure}{0.475\hsize} 
\includegraphics[width=\textwidth]{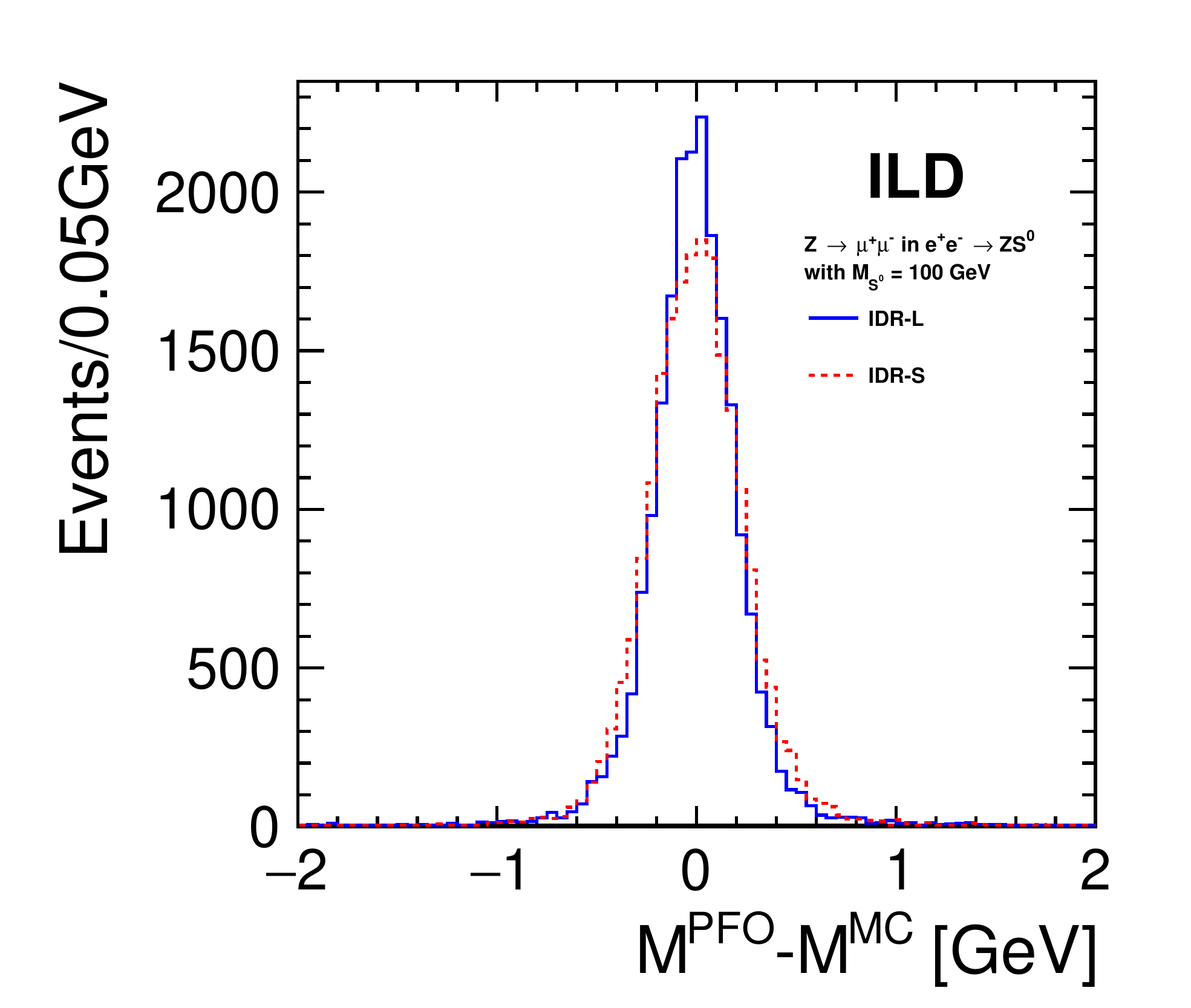}
 \caption{  \label{fig:extraH:Mdiff:mh100}}
 \end{subfigure}
\begin{subfigure}{0.475\hsize} 
\includegraphics[width=\textwidth]{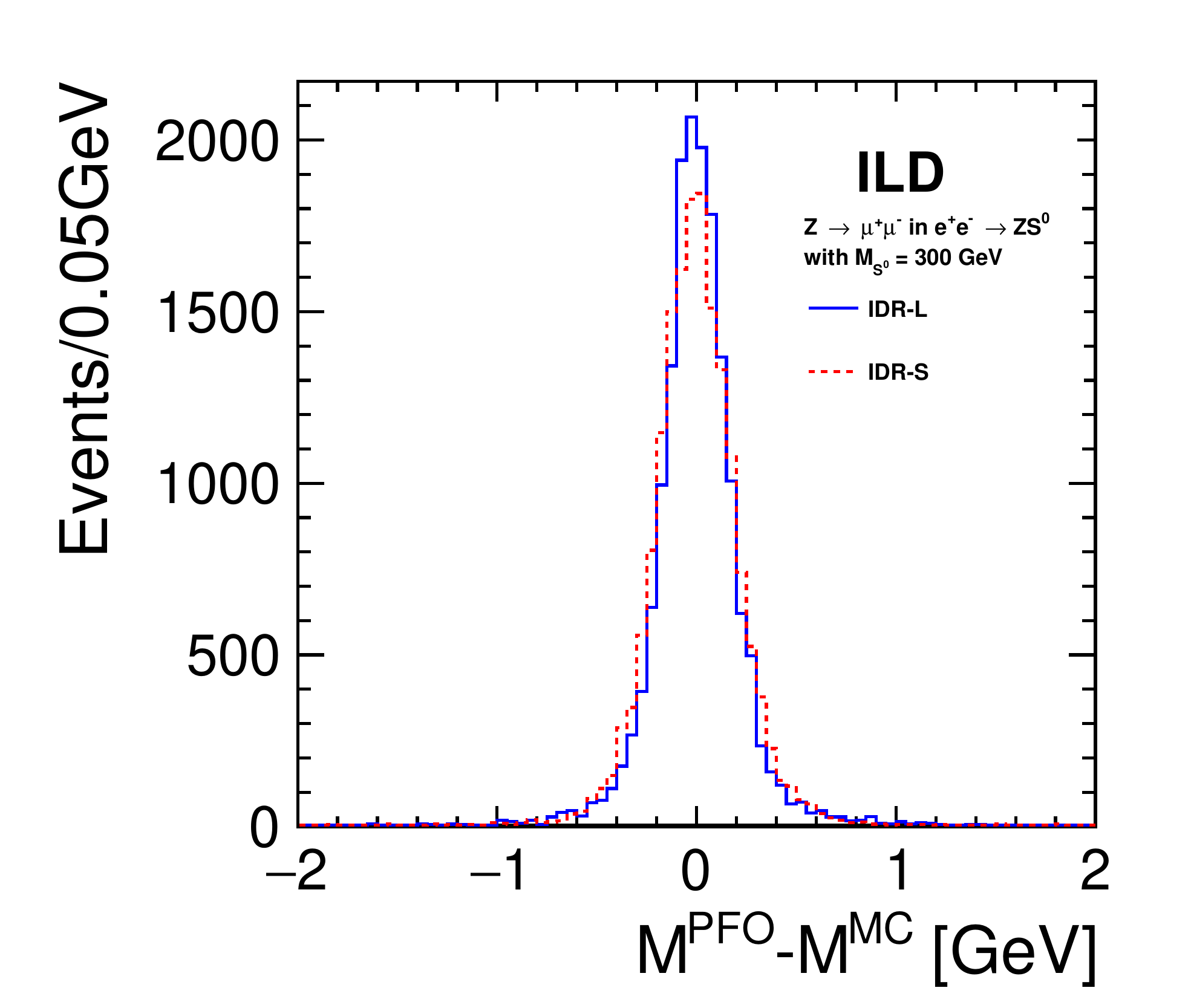}
 \caption{ \label{fig:extraH:Mdiff:mh300}}
 \end{subfigure}
\hspace{0.03\textwidth}
\begin{subfigure}{0.475\hsize} 
\includegraphics[width=\textwidth]{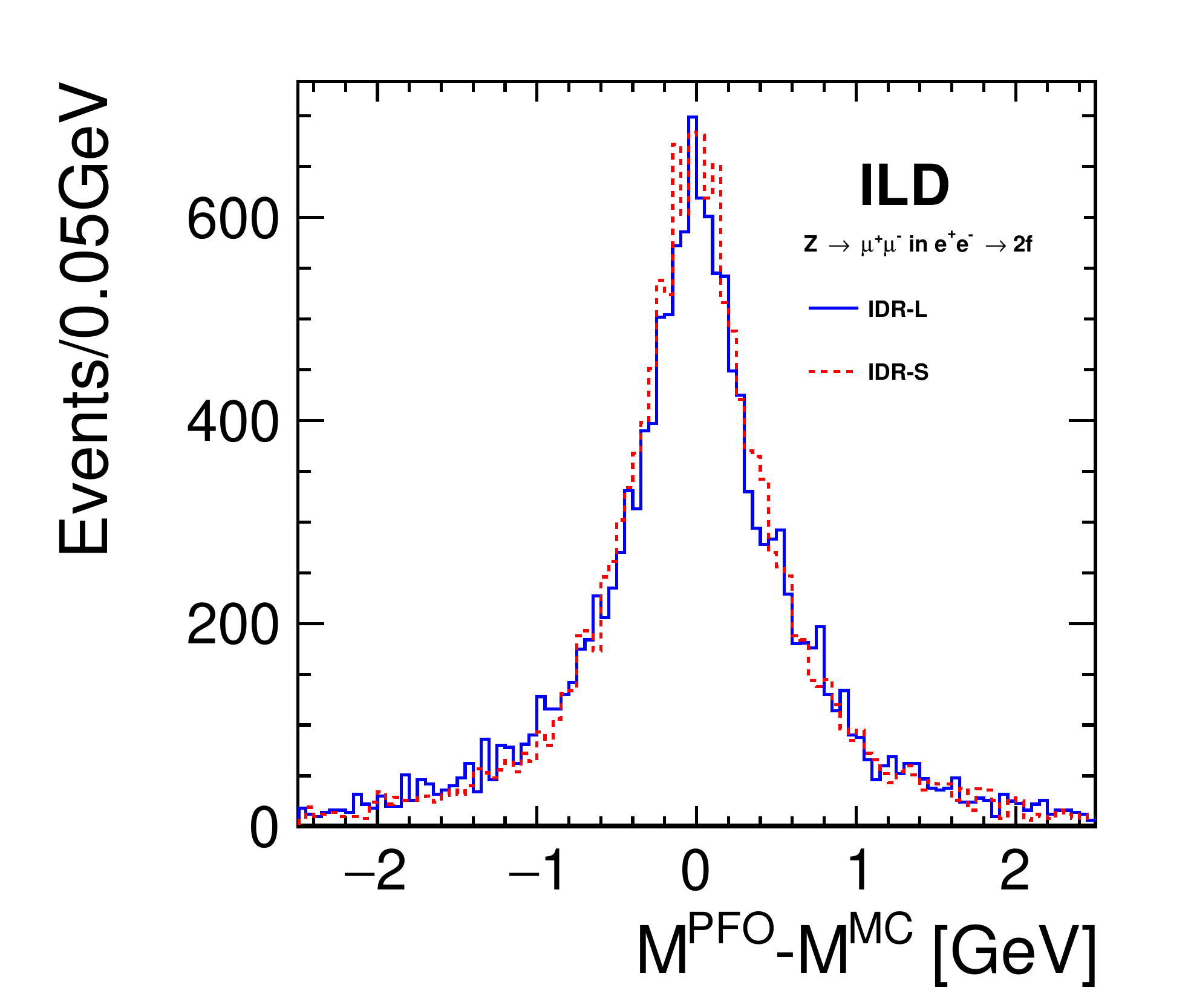}
 \caption{  \label{fig:extraH:Mdiff:2f}}
 \end{subfigure}
\end{center}
\caption{Event-by-event difference between reconstructed and generated di-muon mass for different types of events sampling different ranges in polar angle and momentum of the muons  in IDR-L and IDR-S:
(a) $ZS^0$ signal with $\ms = 20$\,GeV
(b) $ZS^0$ signal with $\ms = 100$\,GeV
(c) $ZS^0$ signal with $\ms = 300$\,GeV
(d) $2f$ background.
}
\label{fig:extraH:Mdiff}
\end{figure}

In order to eliminate the effect of the natural $Z$ width, Fig.~\ref{fig:extraH:Mdiff} compares the event-by-event difference between reconstructed and generated di-muon mass ($M^{PFO}-M^{MC}$) for the same four event samples in IDR-L and IDR-S. Here, clear differences can be seen between the different event samples and the two detector models. In the case of the two-fermion background, the mass reconstruction is worse than for all signal benchmarks, because about half of the events have significant ISR radiation and return to the $Z$ pole. These $Z$ bosons are highly boosted into the forward region, resulting in forward-going muons with a small opening angle. In case of the signal, the $Z$ boson is less boosted when the scalar mass is higher, thus more muons are in the central region of the detector. 
For the two-fermion case, the mass reconstruction in IDR-S and IDR-L turns out about equally precise, since the better performance of IDR-S in the forward region relevant for the radiative-return events is averaged with the better performance of IDR-L for the high-momentum central tracks from the di-fermion events without significant ISR. 
In case of the signal samples the amount of ISR is limited, especially when the scalar mass is higher, and most of the muons are in the barrel region, where IDR-L has better resolution. As a result, the large detector has a better performance for reconstruction of the signal.

\begin{figure}[htbp]
\begin{center}
  \begin{subfigure}{0.475\textwidth}
    \includegraphics[width=\textwidth]{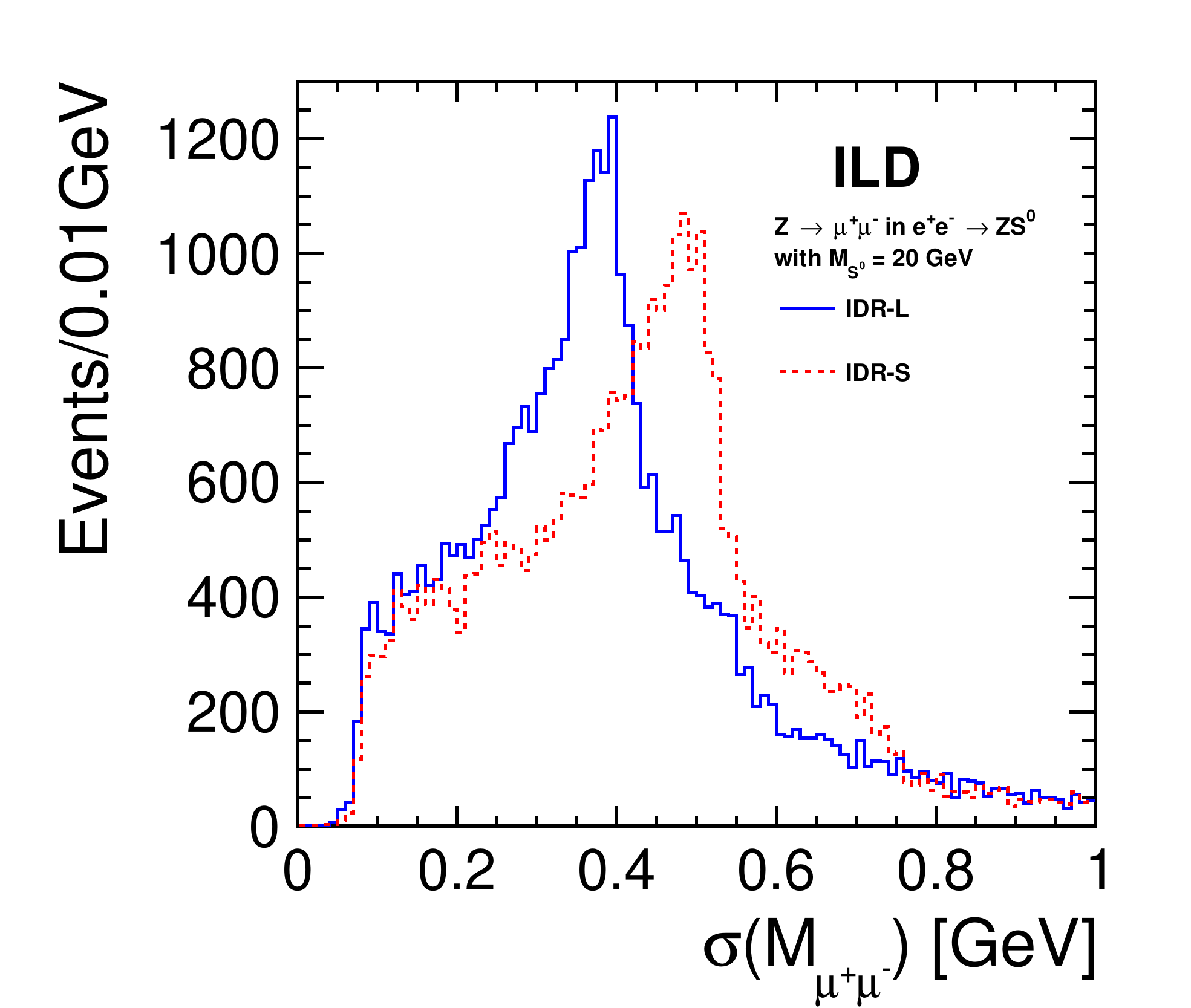} 
    \subcaption{}
  \end{subfigure}  
\hspace{0.03\textwidth}
  \begin{subfigure}{0.475\textwidth}
    \includegraphics[width=\textwidth]{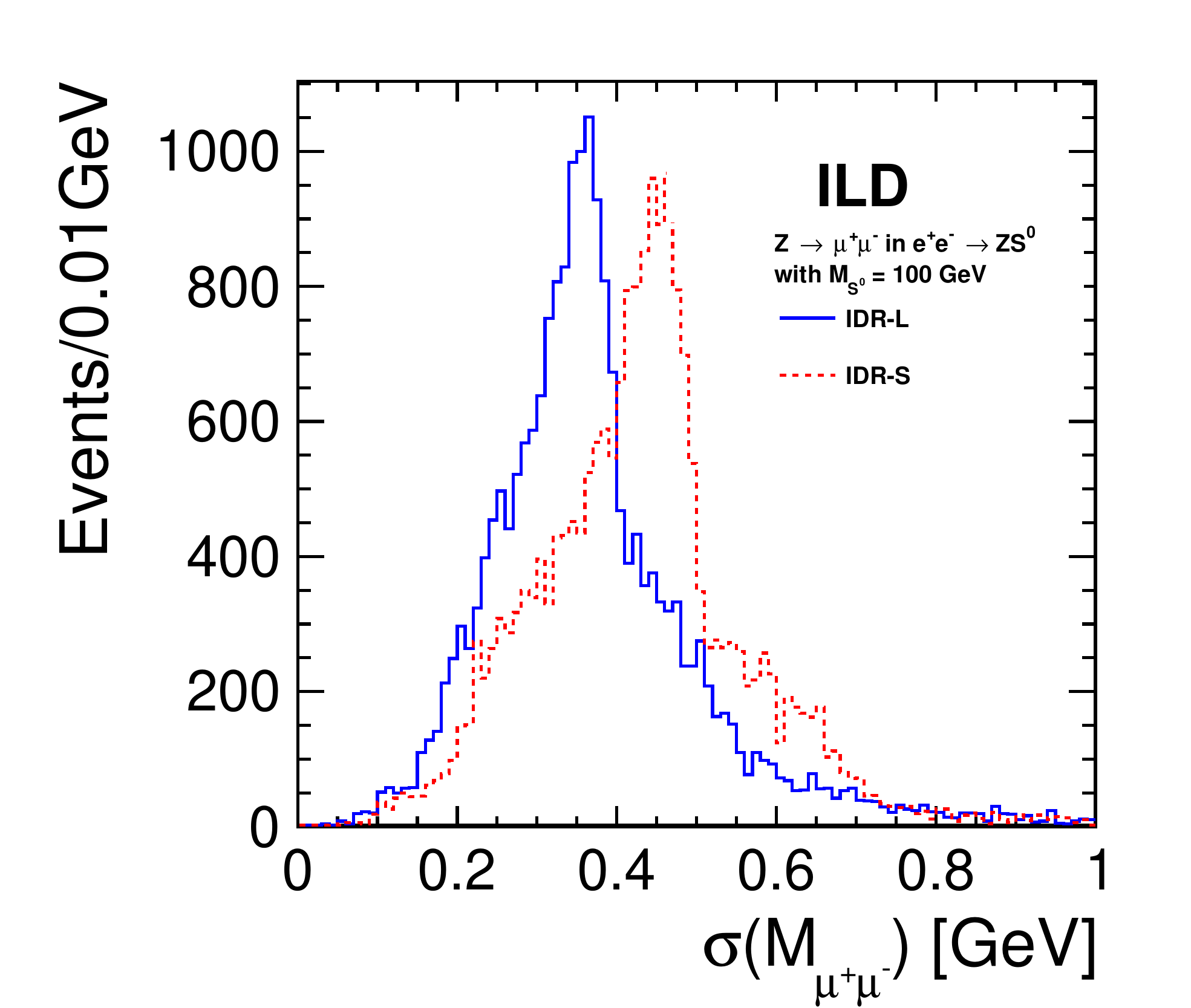} 
    \subcaption{}
  \end{subfigure} 
  \begin{subfigure}{0.475\textwidth}
    \includegraphics[width=\textwidth]{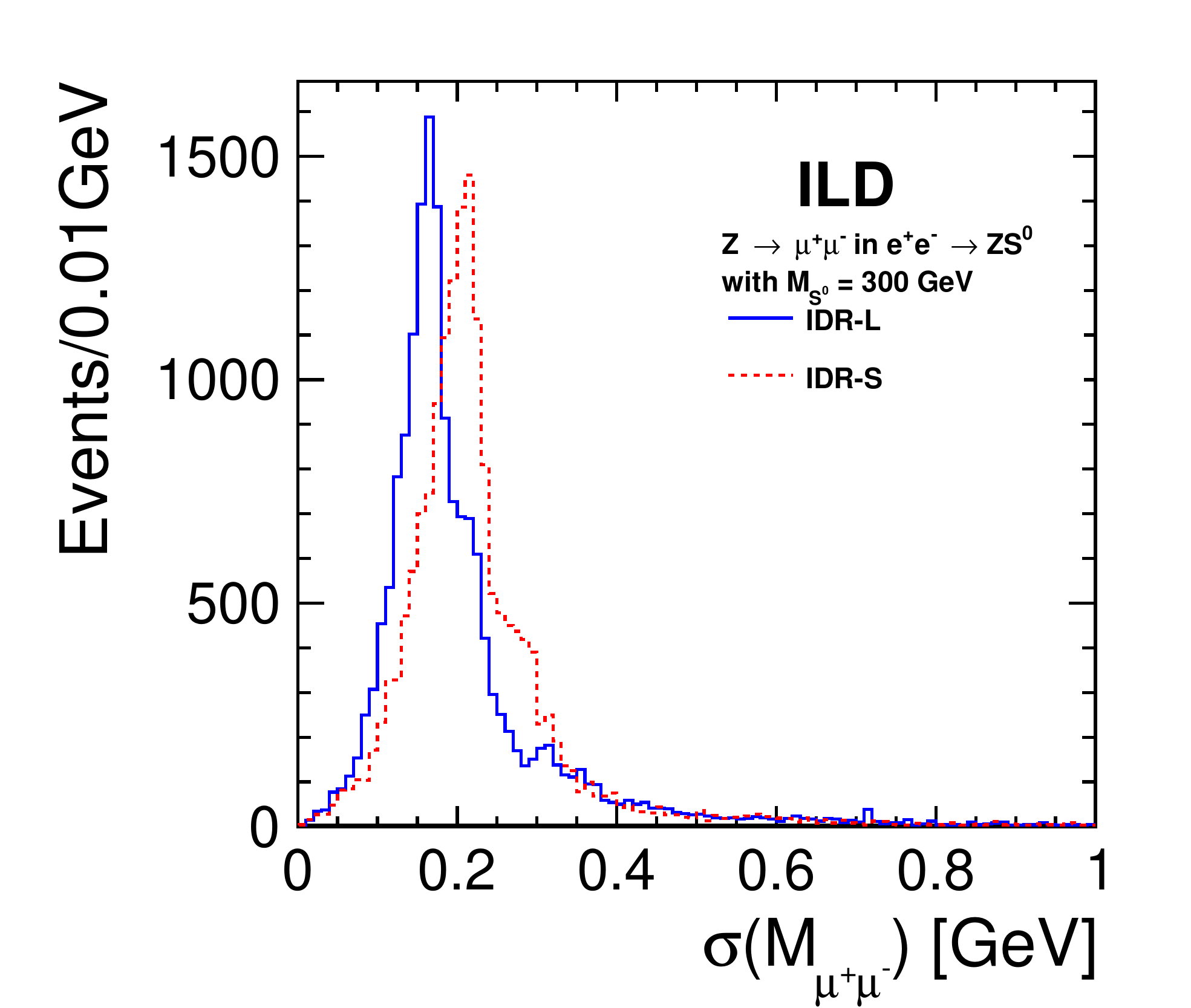} 
    \subcaption{}
  \end{subfigure}
\hspace{0.03\textwidth}
  \begin{subfigure}{0.475\textwidth}
    \includegraphics[width=\textwidth]{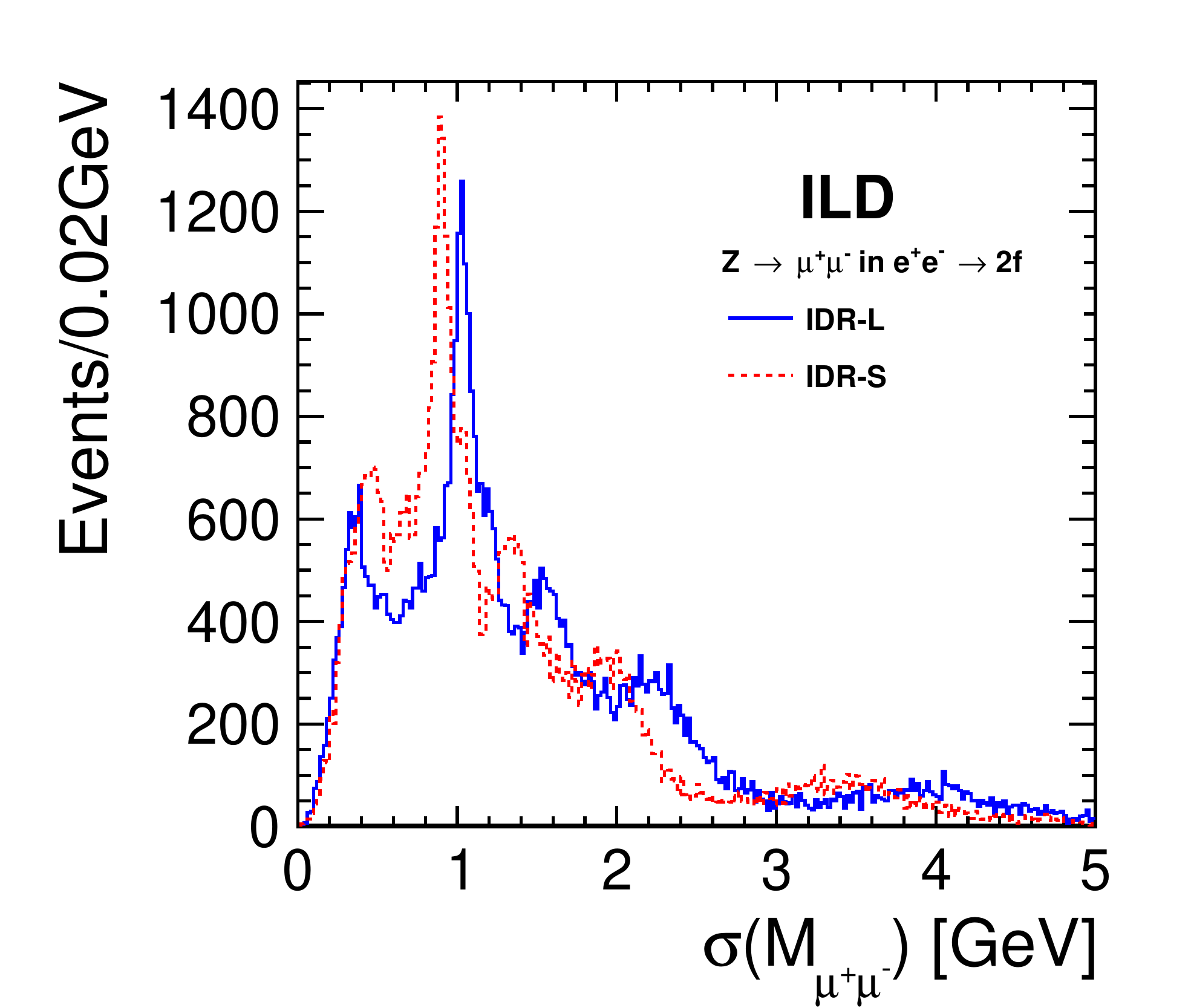} 
    \subcaption{}
  \end{subfigure} 
\end{center}
  \caption{Event-by-event uncertainty on the invariant di-muon mass calculated by error propagation from the track fit for different types of events sampling different ranges in polar angle and momentum of the muons in IDR-L and IDR-S:
(a) $ZS^0$ signal with $\ms = 20$\,GeV
(b) $ZS^0$ signal with $\ms = 100$\,GeV
(c) $ZS^0$ signal with $\ms = 300$\,GeV
(d) $2f$ background.}\label{fig_inm_sigma}
\end{figure}

Finally, Fig.~\ref{fig_inm_sigma} shows the distributions of the event-by-event uncertainty on the di-muon mass, $\sigma_{\mmu}$, propagated from the uncertainties on the track parameters from the track fit. In the case of the signal, the large detector always has a better resolution. In the case of the two fermion background, in the small $\sigma_{\mmu}$  region, the uncertainties are similar to the signal, where the muons are mostly in the central region. When $\sigma_{\mmu}>0.6$ GeV, which typically occurs in the forward region, the small detector performs better than the large detector.

Allthough there are some clear differences between the two detector models visible in the di-muon reconstruction, these will not propagate in any significant way to the final result, since in the recoil mass spectrum, they will be covered by the $Z$ width and the beam energy spectrum.

\subsection{Identification of ISR Photons}
\label{subsec:ISR}

For the identification of ISR photons which reach the detector, an algorithm has been developed in the context of this analysis based on a gradient boosted decision tree (BDTG) as implemented in \textsc{TMVA}~\cite{MVA} in \textsc{ROOT}~\cite{Brun:1997pa}. All physics processes which
are often accompanied by significant ISR are included in the training, i.e.\ SM Higgs production and the two- and four-fermion processes. All photon candidates with an energy above $10$\,GeV ($5$\,GeV) for $|\cos{\theta}|<0.95$ ($|\cos\theta|>0.95$) are considered. Real ISR photons are identified by the MC truth form the ``signal'', while all the other photons, including FSR and bremsstrahlung are treated as ``background''.  
A double-cone method, similar to the $\textsc{IsolatedLeptonTagging}$ processor described in Sec.~\ref{subsec:presel}, is employed to quantify the degree of isolation. For each photon candidate, small and large cones around the photon are constructed with opening angles of $\cos{\theta_s} = 0.98$ and $\cos{\theta_l} = 0.95$. Figure~\ref{fig_photon_MVA_input} shows the ten observables used as input to the BDTG, as well as the distribution of the BDTG output:

\begin{figure}[htb]
 \begin{center}	
 \includegraphics[width=0.95\textwidth]{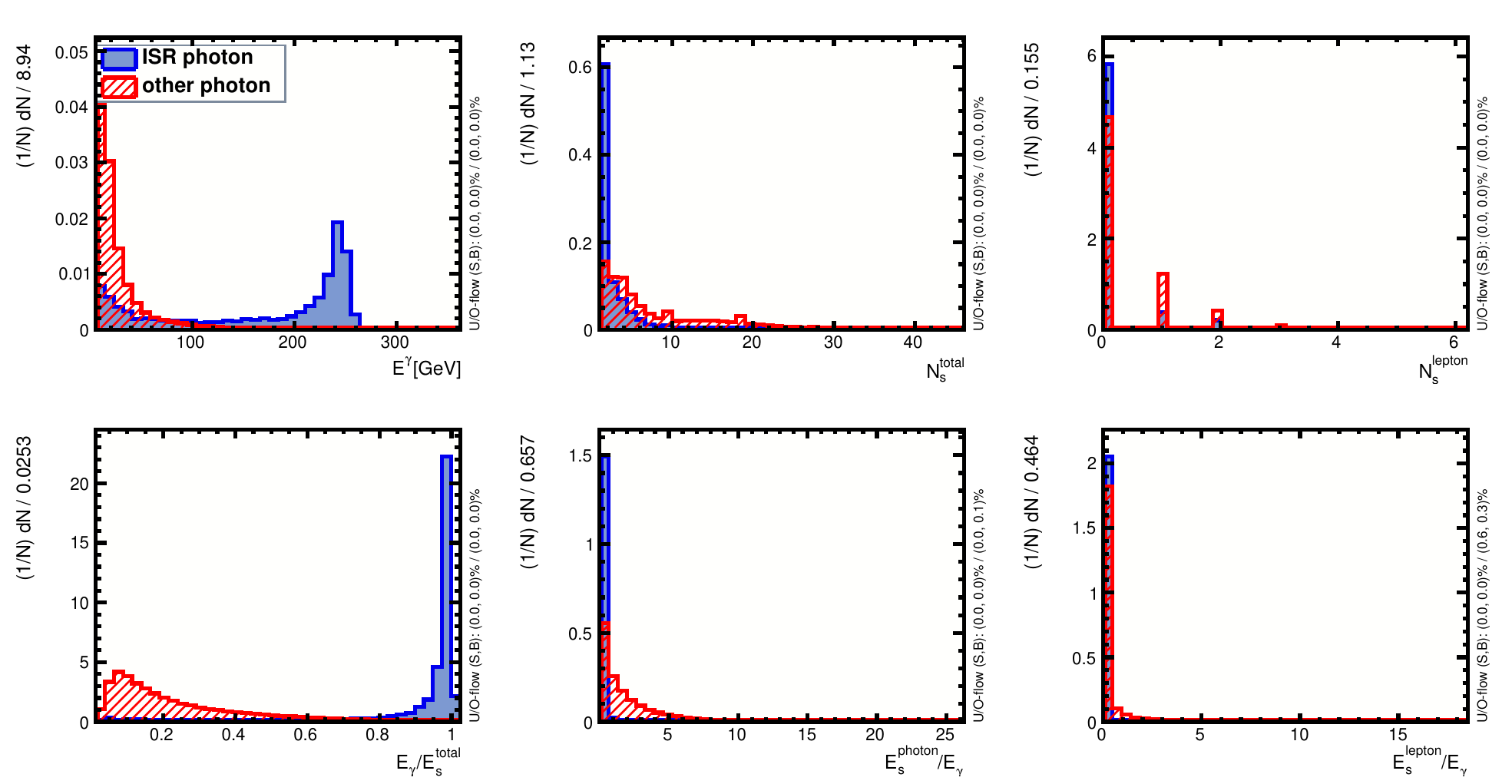} 
 \includegraphics[width=0.95\textwidth]{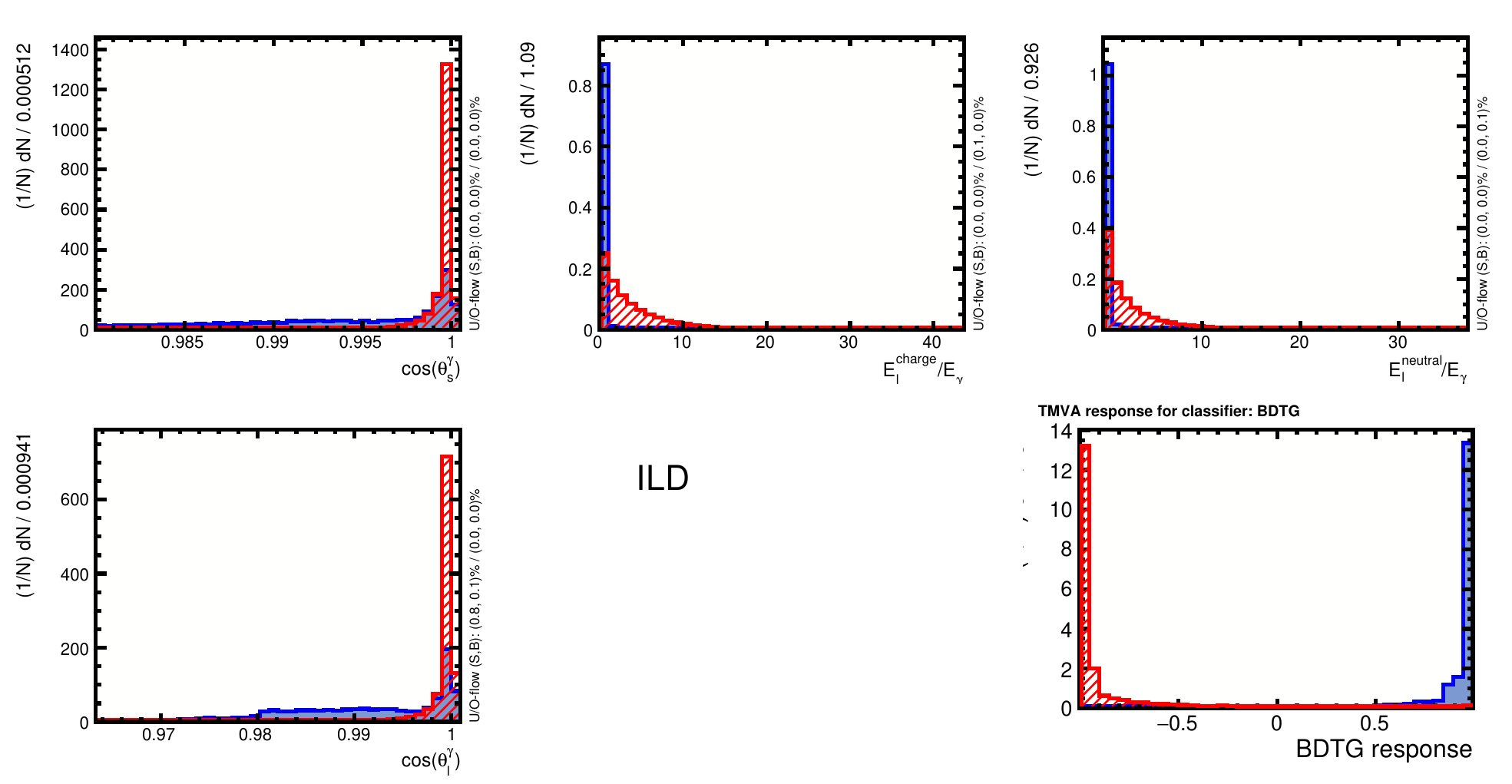} 
 \end{center}
 \caption{The input observables and the BDTG output for the ISR photon identification.}\label{fig_photon_MVA_input}
\end{figure}

\begin{itemize}
\item the candidate photon's energy $E^{\gamma}$
\item the total number of particles in the small cone  $N_{s}^{\mathrm{total}}$
\item the number of leptons in the small cone $N_{s}^{\mathrm{lepton}}$

\item the ratio of the candidate photon's energy over the total energy sum in the small cone $E_{\gamma}/E_{s}^{\mathrm{total}}$
\item the ratio of the energy sum of all other photons in the small cone over the candidate photon's energy  $E_{s}^{\mathrm{photon}}/E_{\gamma}$ 
\item the ratio of the energy sum of all leptons in the small cone over the candidate photon's energy  $E_{s}^{\mathrm{lepton}}/E_{\gamma}$
\item the ratio of the energy sum of all charged particles in the large cone over the candidate photon's energy $E_{l}^{\mathrm{charge}}/E_{\gamma}$
\item the ratio of the energy sum of all other neutral particles in the large cone over the candidate photon's energy $E_{l}^{\mathrm{neutral}}/E_{\gamma}$
\item the cosine of the opening angle between the candidate photon and the combined four-momentum of all other particles in the small cone $\cos{\theta_{s}^{\gamma}}$ 
\item the cosine of the opening angle between the candidate photon and the combined four-momentum of all other particles in the large cone $\cos{\theta_{l}^{\gamma}}$. 
\end{itemize} 

\begin{figure}[htb]
 \begin{center}		
   \includegraphics[width=0.6\textwidth]{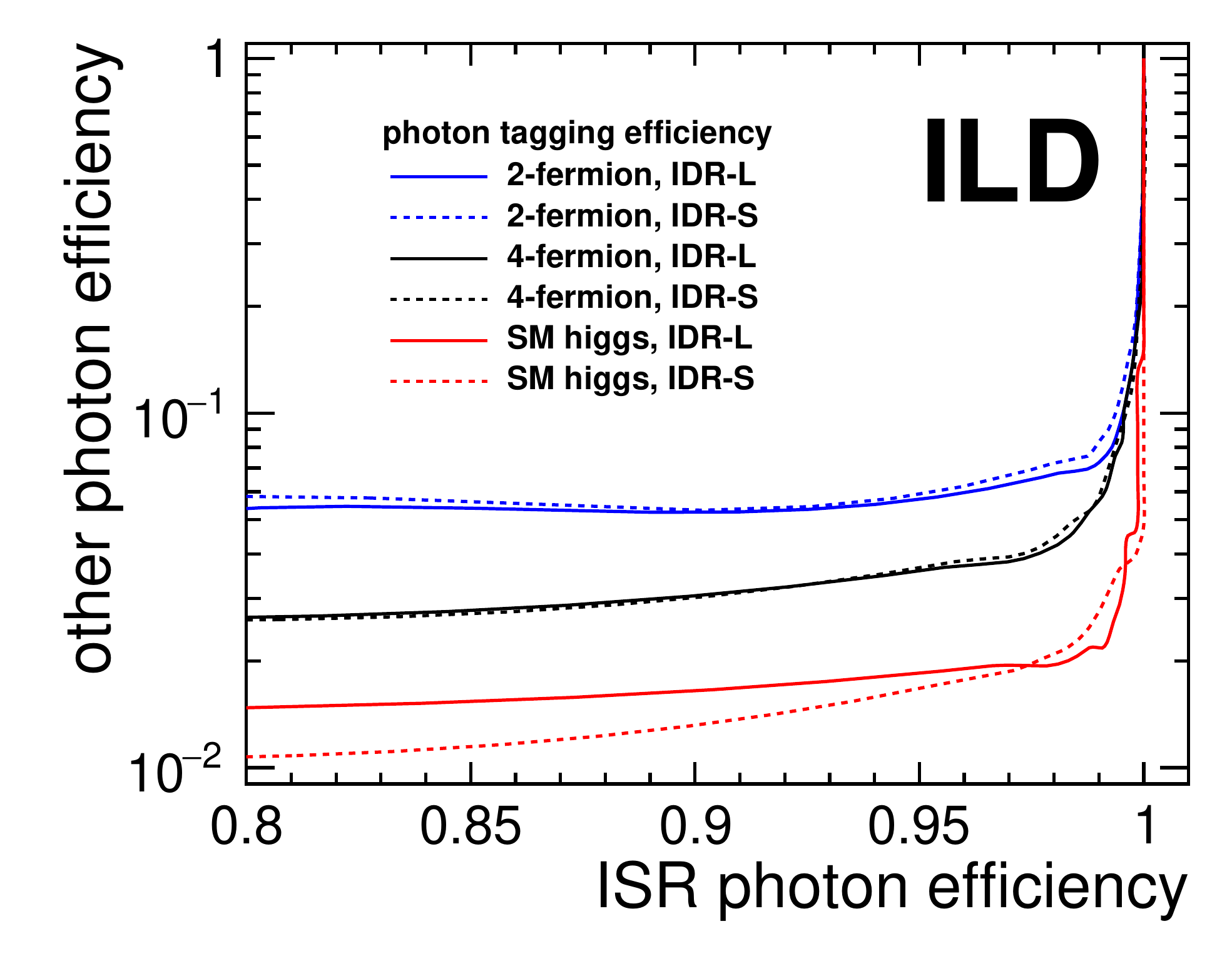} 
 \end{center}
 \caption{The ISR photon tagging efficiency for the ISR photons and non-ISR photons in IDR-L and IDR-S. The blue/black/red curves are the efficiencies for two-fermion/four-fermion/higgs samples, respectively.}\label{fig_isr}
\end{figure}

Figure~\ref{fig_isr} shows the mistagging rate as a function of the ISR tagging efficiency in the IDR-L and IDR-S models and for the two-fermion, four-fermion and SM-higgs-boson samples. The ISR photon tagging performs very well. The BDTG output is required to be larger than $0.8$, which in case of the two-fermion samples corresponds to an efficiency of $99.3\%$ at a mistagging rate for other photons of $8.1\%$. For tighter cuts on the BDTG output, the mistagging rate remains nearly constant, while only the signal efficiency drops. Due to the large cross section of the two-fermion processes, and the large number of cases where the ISR photon escapes undetected, there will be still a large number of events with a non-ISR photon mistagged as ISR. Both detector models perform very similar in terms of the ISR identification.

The ISR information will be used in several ways. If an ISR photon is identified in an event, this photon will be excluded from the bremsstrahlung/FSR recovery described in the next section, and the kinematics of events with an ISR photon will be corrected by boosting the rest of the event, in particular the muons (incl.\ any recovered bremsstrahlung/FSR photons), into the reduced centre-of-mass frame of the hard interaction. These boosted quantities will be used in the MVAs described in Sec.~\ref{subsec:MVA}. Since in particular the radiative return events have much higher ISR photon energies than possible in $ZS$ production, events with a very high-energetic ISR photon will be vetoed in the last analysis step as described in Sec.~\ref{subsec:select}.

\subsection{Recovery of Bremsstrahlung and FSR Photons}
\label{subsec:FSR}
After selecting the muon pair (c.f.\ Sec.~\ref{subsec:presel}) and possibly an ISR candidate (c.f.\ Sec.~\ref{subsec:ISR}), bremsstrahlung and final-state radiation (FSR) photons are identified and recombined with the muons.
The four-momenta of photons which fulfill $\cos\theta_{\gamma-\mu} > 0.99$ are added to the four-momentum of the muon, following the procedure developed for the SM Higgs recoil analysis~\cite{Yan:2016xyx}.
%The bremsstrahlung and FSR are identified using its polar angle with respect to the final state muons. If the cosine of the polar angle is larger than 0.99, the photon four momentum is combined with the muon momentum.
 
\begin{figure}[htb]
      \begin{center}
\includegraphics[width=0.6\textwidth]{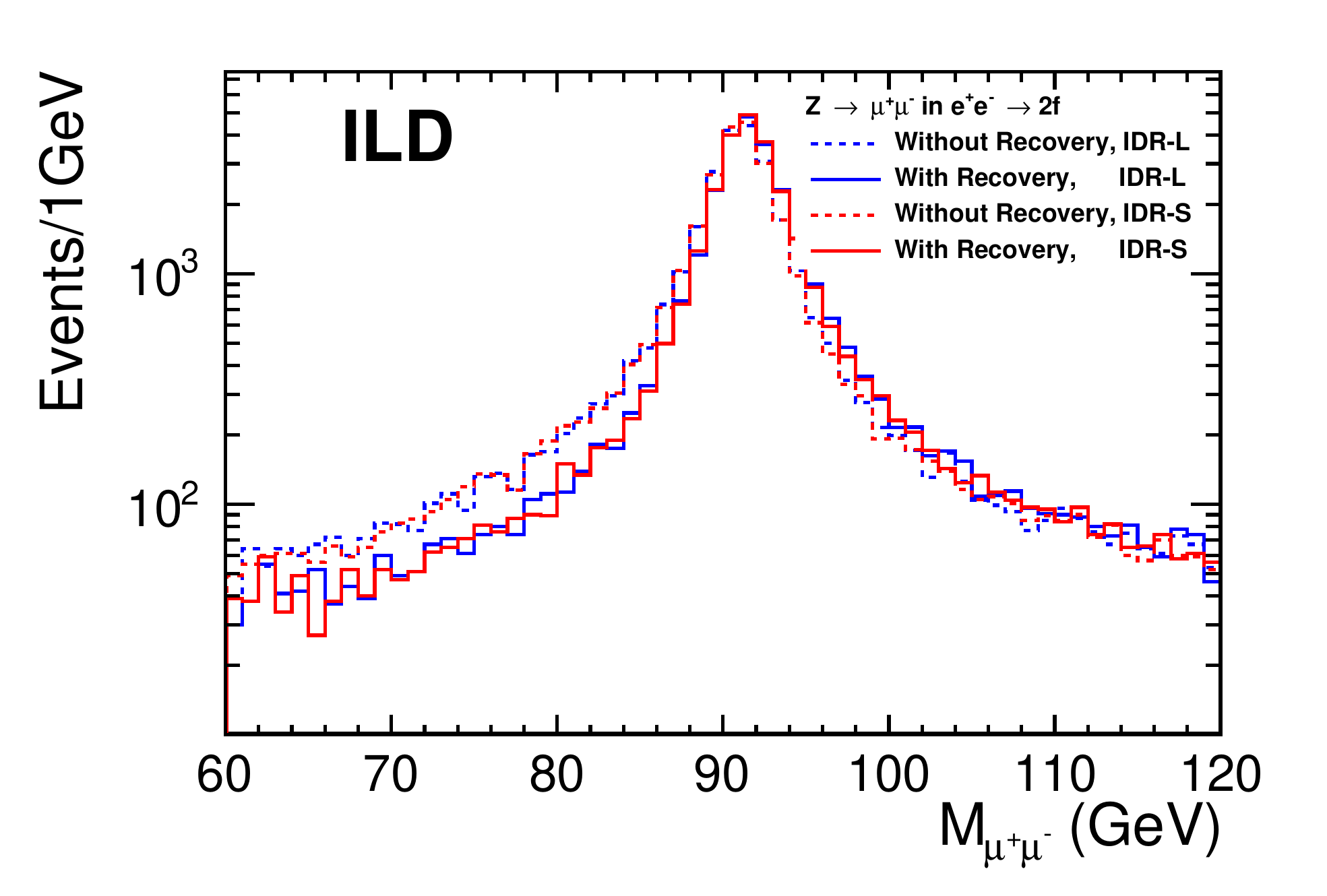} 
      \end{center}
  \caption{Comparison of the reconstructed $Z$ candidate mass in the two-fermion samples with (solid lines) and without (dashed lines) bremsstrahlung/FSR recovery for IDR-L (blue) and IDR-S (red).}\label{fig_recovery}
\end{figure}

Figure \ref{fig_recovery} compares the reconstructed $Z$ candidate mass distribution in the two-fermion samples before and after bremsstrahlung/FSR recovery for IDR-L and IDR-S. The dashed lines show the invariant mass of the pure muons without a bremsstrahlung/FSR  recovery, and are the same as in Fig.~\ref{fig_inm:2f}, while the solid lines include bremsstrahlung/FSR recovery, which will be used in the further selections (e.g.\ Fig.~\ref{fig_2f_MVA_input}(a) and~\ref{fig_4f_MVA_input}(a)). It can be seen that the recovery process concentrates the $M_{\mu^{+}\mu^{-}}^{FSR}$ spectrum to the peak, especially for the events in the lower end tail. The peak position is shifted to higher masses by only about $60$\,MeV by the photon recovery procedure. Still, the post-recovery distribution seems to have a somewhat over-compensating effect. Here, there is still room for improvement by developing a more sophisticated algorithm. With the current reconstruction, the difference between IDR-L and IDR-S is much smaller than the effect of the bremsstrahlung/FSR recovery.

\subsection{Rejection of the Two- and Four-Fermion Backgrounds}
\label{subsec:MVA}
The main selection comprises cuts on the output of two BDTGs, one trained specifically against the two-fermion background (2f-MVA), and the other against all other backgrounds, which are dominated by the four-fermion events (4f-MVA). Both BDTGs are trained for each scalar mass. The two-fermion processes are the most important background for low mass scalars, while the four-fermion processes are dominant when the scalar mass is above the $Z$ resonance. Six-fermion backgrounds only become relevant for scalar masses above $200$\,GeV. As a result, training two different BDTGs not only helps the optimization of the BDTGs themselvs, but also allows to efficienctly adjust the selection for the different ranges of scalar masses by choosing appropriate cuts on the BDTG outputs.
Both BDTGs are trained using the same six input variables, all calculated after the FSR recovery, and after boosting into the centre-of-mass system of the hard interaction based on the highest-energetic ISR candidate found in the event: 
\begin{itemize}
\item the mass of the $Z$ boson candidate: $M_{\mu^{+}\mu^{-}}^{FSR}$,
\item the polar angle of each muon: $\textrm{cos}\theta_{\mu^{+}}^{FSR}$,  $\textrm{cos}\theta_{\mu^{-}}^{FSR}$,
\item the polar angle of the $Z$ boson candidate: $\textrm{cos}\theta_{\mu\mu}^{FSR}$,
\item the opening angle of the muon pair: $\textrm{cos}\theta_{\mu-\mu}^{FSR}$,
\item the acoplanarity between the muons: $\pi-(\phi_{\mu^{+}}-\phi_{\mu^{-}})$.
\end{itemize}
There are also some other variables, which may reflect the difference between signal and backgrounds, such as $\sigma_{M_{\mu^{+}\mu^{-}}}$, as shown in Fig.~\ref{fig_inm_sigma}. However, since the mass resolution mainly depends on the polar angle of the muons, this 
variable is highly correlated with the inputs listed above and does not improve the BDTG when added as an additional input.

\begin{figure}[htbp]
  \begin{center}	
    \includegraphics[width=\textwidth]{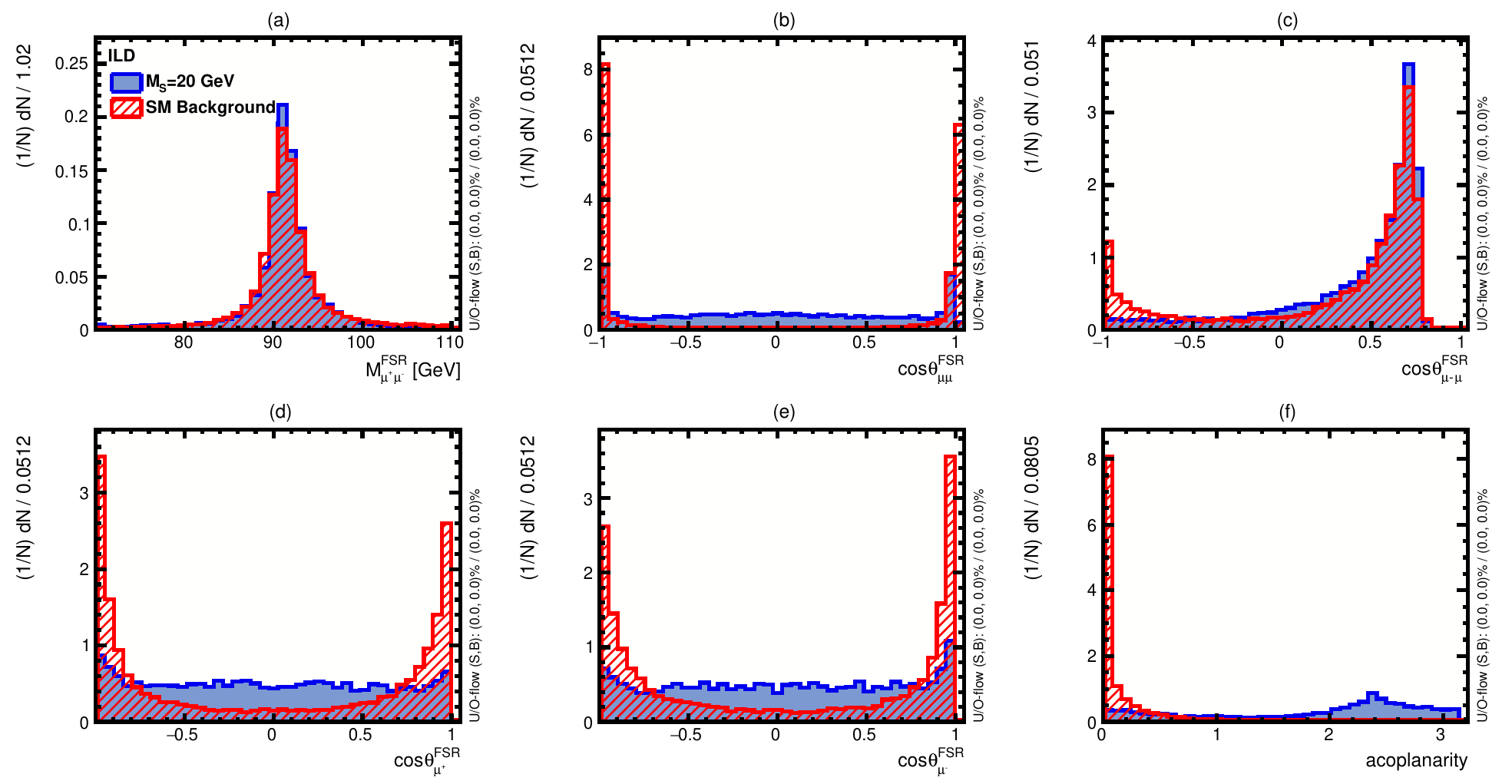} 
  \end{center}
  \caption{The input distributions to the 2f-MVA for $\ms=20$\,GeV based on IDR-L.}\label{fig_2f_MVA_input}
\end{figure}

\begin{figure}[htbp]
    \begin{center}	
 \includegraphics[width=\textwidth]{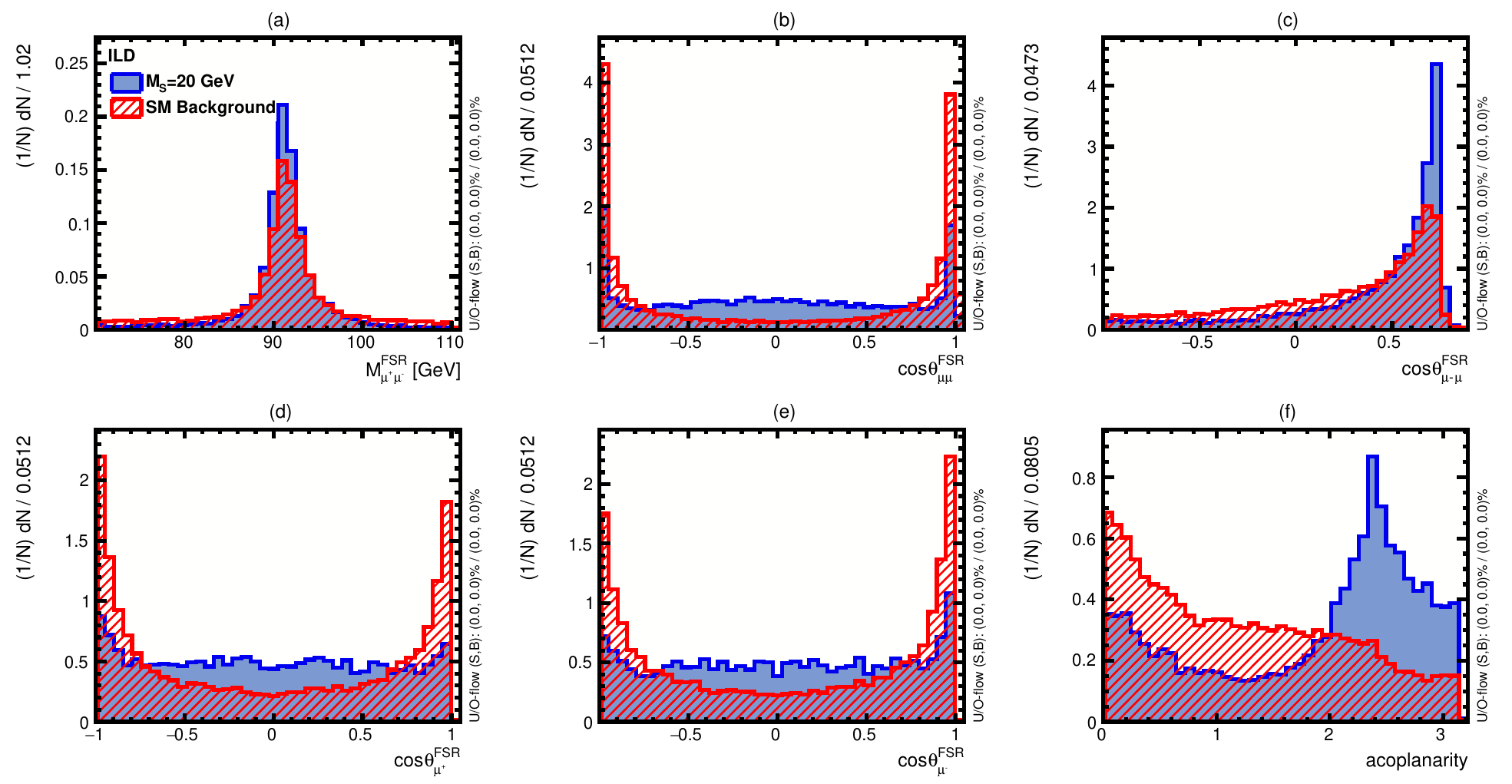} 
  \end{center}
  \caption{The input distributions to the 4f-MVA for $\ms=20$\,GeV based on IDR-L.}\label{fig_4f_MVA_input}
\end{figure}

\begin{figure}[htbp]
\begin{center}
  \begin{subfigure}{0.475\textwidth}
    \includegraphics[width=\textwidth]{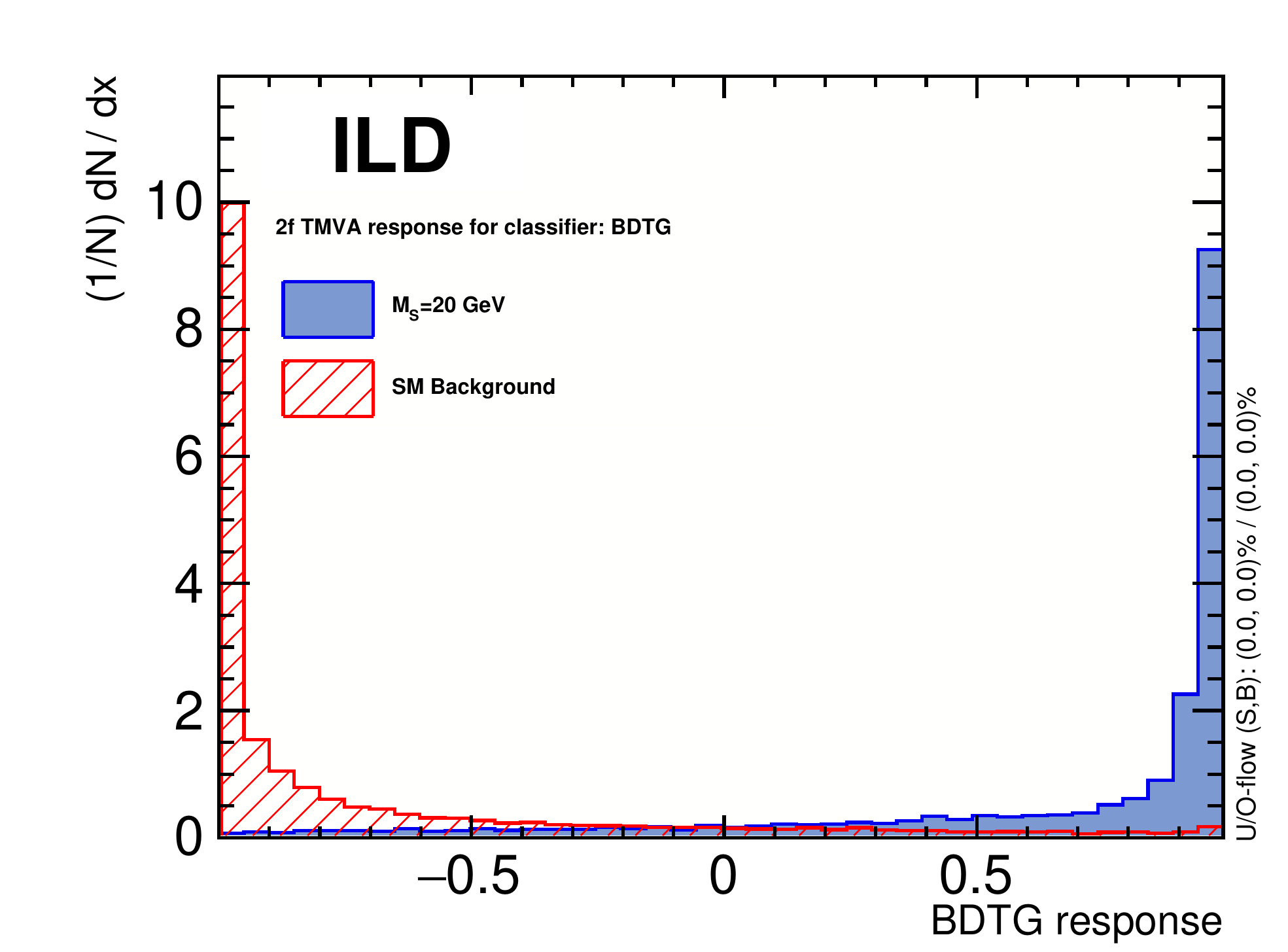} 
    \subcaption{\label{fig:bdtgout:2f}}
  \end{subfigure}  
\hspace{0.03\textwidth}
  \begin{subfigure}{0.475\textwidth}
    \includegraphics[width=\textwidth]{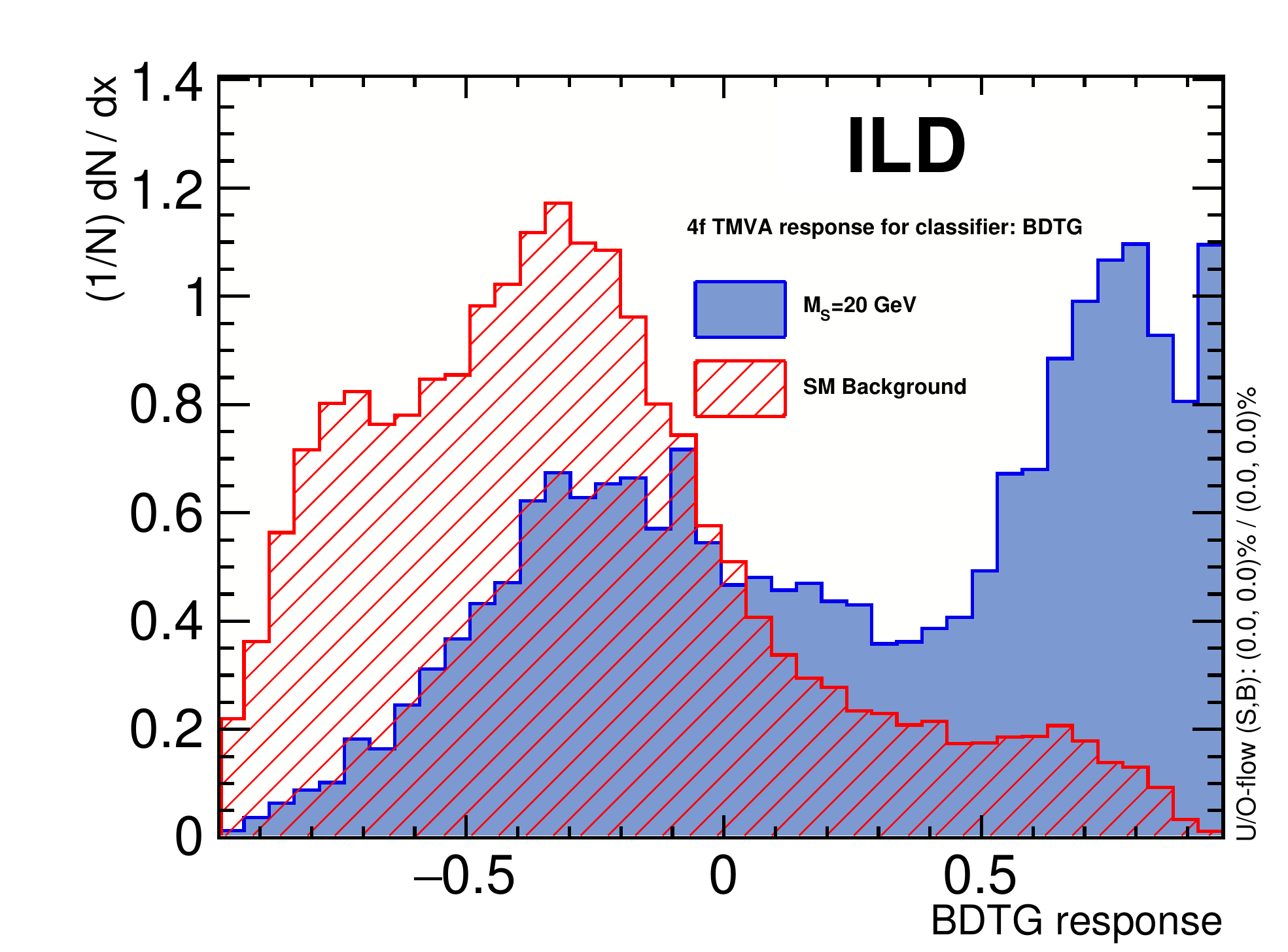} 
    \subcaption{\label{fig:bdtgout:4f}}
  \end{subfigure} 
\end{center}
  \caption{Output distributions of the BDTGs trained for the case of $\ms = 20$\,GeV based on IDR-L:
(a) 2f-MVA
(b) 4f-MVA.
The corresponding input distributions are shown in Figs.~\ref{fig_2f_MVA_input} and~\ref{fig_4f_MVA_input}, respectively. }\label{fig:bdtgout}
\end{figure}

Figures~\ref{fig_2f_MVA_input} and~\ref{fig_4f_MVA_input} show the input variable distributions obtained with IDR-L for the 2f-MVA and the 4f-MVA, respectively, for the example of $\ms=20$\,GeV.  Figure~\ref{fig:bdtgout} displays the corresponding BDTG outputs. The 2f-MVA gives a very clear separation between the signal and the 2f backgrounds, while there is large overlap area between the signal and backgrounds in 4f-MVA output. This difference arises to a large extent due to the much better separation power of the acoplanarity in the 2f-MVA. As can be seen in Fig.~\ref{fig_2f_MVA_input}(f), the two muons are almost back-to-back (acoplanarity $\simeq 0$) in case of the two-fermion background, especially since in cases where the ISR photon is detected the event is boosted into the centre-of-mass frame of the hard interaction. In case of the four-fermion background, the acoplanarity distribution has a much larger overlap with the signal, c.f.\ Fig.~\ref{fig_4f_MVA_input}(f). 

\subsection{Main Selection}
\label{subsec:select}\
The ingredients presented in the previous sections are then brought together in the actual main event 
selection. Following the loose preselection of $Z\to\mu^+\mu^-$ candidates introduced in 
Sec.~\ref{subsec:presel}, cuts on the variables listed in Table~\ref{table_cut_content} 
are applied. A full optimisation of the cut values has been performed for the IDR-L model, and after several cross checks it turned out that  the same values can safely be employed for the IDR-S analyses.  
%The $M_{\mu^{+}\mu^{-}}^{FSR}$, $P^{T,FSR}_{\mu^{+}\mu^{-}}$, $\sigma(M_{\mu^{+}\mu^{-}})$ and ISR photon veto cut values are always the same for different scalar masses of both IDR-L and IDR-S. The MVA-related cut values will be tuned a little which is summarized in Appendix.
The ISR photon cut vetoes events with very high energetic ISR photons, in particular 
two-fermion events returning to the $Z$ pole, which corresponds to a photon energy of 
about $240$\,GeV.  Considering the reconstruction efficiency in different angular 
regions of the detector, events with any identified ISR photon candidate with $E_{\gamma}>230$\,GeV for $|\cos{\theta}|<0.95$ or $E_{\gamma}>150$\,GeV for $|\cos{\theta}|>0.95$ are vetoed.

\begin{table}[htb]
 \begin{center}
 \begin{small}
 \begin{tabular}{|c| c|}
 \hline
 $      M_{\mu^{+}\mu^{-}}^{FSR}$ & $\in [ 70 ,110 ] $\,GeV\\
 \hline
 $P^{T,FSR}_{\mu^{+}\mu^{-}}$ & $\in [ 0 ,245 ] $\,GeV\\ 
 \hline
 $    \sigma(M_{\mu^{+}\mu^{-}})$&$ \in [ 0 ,1 ] $\,GeV\\ 
 \hline
 $                    MVA_{2f} $&$ \in  [ 0.75 ,1 ] $ \\ 
 \hline
 $                      MVA_{4f} $&$ \in  [ 0.5 ,1 ] $ \\ 
 \hline
 ISR photon &  none, or $   E_{\gamma}^{\textrm{central(forward)}} < 230 (150) $\,GeV  \\ 
 \hline
% $                  M_{\textrm{rec}} $&$\in [ 0, 450 ] $ GeV  \\ 
% \hline
 \end{tabular}
 \end{small}
 \end{center}

  \caption{The cuts after the preselection for IDR-L and IDR-S. The MVA-cuts have been optimised 
  for each scalar mass, the values given here correspond to $\ms = 20$\,GeV. The corresponding 
  values for all signal masses can be found in Table~\ref{table_cut_masses} in the Appendix.}
  \label{table_cut_content}
 \end{table}

The optimization of the cut values has been performed based on the significance 
$\Sigma = S/\sqrt{S+B}$, where $B$ is the number of remaining background events, 
and $S$ is the number of signal events expected if the scalar had the same cross 
section as predicted by the SM for a Higgs boson with the same mass. Thereby, both 
$S$ and $B$ are counted in a mass window of $\ms - 20$\,GeV$ < \mrec < 450$\,GeV. The whole 
event selection does not make use of any property of the decay products of the extra scalar.

Examples of the cut flow for IDR-L and IDR-S are given for the $P(e^-,e^+)=(-80\%,+30\%)$ 
data set, normalised to its nominal luminosity of $1600$\,fb$^{-1}$, for the cases 
of $\ms=20$\,GeV, $\ms=100$\,GeV and $\ms=300$\,GeV in Table~\ref{table_cutflow}. 
The last row in each table gives the event numbers in the mass window
used for the cut optimisation. This is, however, for information only, since the full mass range, i.e.\ all events remaining after the ISR veto, will be passed to the final sensitivity calculation.
All background processes which have a sizable event count after the $       M_{\mu^{+}\mu^{-}}^{FSR}  $ \& $   P^{T,FSR}_{\mu^{+}\mu^{-}}$ cut are shown in individual columns. The column $B_{\mathrm{total}}$ includes all analysed background processes.

 \begin{table}[p]
 \vspace{-0.5cm}
 \begin{center}
\subcaption{ IDR-L, $\ms=20$ GeV}
 \begin{scriptsize}
 \begin{tabular}{|c| c| c| c| c| c| c| c| c| c|}
 \hline
   &  $             \ms=20 $  &  $              \mu^+\mu^- h $ &  $              4f_{l} $ &  $             4f_{sl} $ &  $              2f_{l} $ &  $            6f_{all} $ &  $B_{\mathrm{total}}$ &               $\epsilon$ &             $\Sigma$\\ 
\hline
no cut ($ \sigma\times \int\mathcal{L}dt$) &                  10459.6 &                  5519.2 &                1.69$\times 10^{7}$ &                  2.11$\times 10^{7}$ &                5.44$\times 10^{6}$&                 1.88$\times 10^{6}$ &                   9.06$\times 10^{7}$   &                        1 &        1.10          \\    
 \hline
$       M_{\mu^{+}\mu^{-}}^{FSR}  $ \& $   P^{T,FSR}_{\mu^{+}\mu^{-}}$ &                  7399.8 &                  4805.03 &                   101307 &                  56558.1 &                   539243 &                  5662.55 &                   707576 &                        0.71 &                  8.75\\ 
 \hline
$   \sigma(M_{\mu^{+}\mu^{-}})$ &                  6844.3 &                  4696.5 &                  84630.5 &                  46201.9 &                   246899 &                  5505.6 &                   387933 &                 0.65 &                  10.89\\ 
 \hline
  $                    MVA_{2f} $ &                     4462 &                   4072.9 &                  17488.4 &                  19326.2 &                   8052.9 &                     1808 &                  50748.4 &                      0.43 &                    18.99\\ 
 \hline
 $                    MVA_{4f} $ &                   2908.7 &                   2347.2 &                   3601.8 &                   5674.5 &                   3337.1 &                    177.9 &                  15138.6 &                     0.28 &                    21.65\\ 
 \hline
ISR photon veto &                   2906.7 &                   2347.2 &                   3601.6 &                   5674.5 &                   1058.4 &                    177.9 &                  12859.7 &                     0.28 &                    23.15\\ 
 \hline\hline
 $\ms - 20$\,GeV$ < \mrec  < 450$\,GeV &                   2906.7 &                   2347.2 &                   3601.6 &                   5674.5 &                   1058.4 &                    177.9 &                  12859.7 &                     0.28 &                    23.15\\ 
 \hline
 \end{tabular}
 \end{scriptsize}

\subcaption{  IDR-S, $\ms=20$ GeV}
 \begin{scriptsize}
 \begin{tabular}{|c| c| c| c| c| c| c| c| c| c|}
 \hline
   &  $             \ms=20 $  &  $              \mu^+\mu^- h $ &  $              4f_{l} $ &  $             4f_{sl} $ &  $              2f_{l} $ &  $            6f$ &  $B_{\mathrm{total}}$ &               $\epsilon$ &             $\Sigma$\\ 
 \hline
no cut ($ \sigma\times \int\mathcal{L}dt$) &                  10459.6 &                  5519.2 &                1.69$\times 10^{7}$ &                  2.11$\times 10^{7}$ &                5.44$\times 10^{6}$&                 1.88$\times 10^{6}$ &                   9.06$\times 10^{7}$   &                        1 &        1.10          \\  
 \hline
$       M_{\mu^{+}\mu^{-}}^{FSR}  $ \& $   P^{T,FSR}_{\mu^{+}\mu^{-}} $&       7394.4 &                  4804.5 &                   102839 &                  56861.5 &                   540588 &                  6150.0 &                   711243 &                        0.71 &                  8.72\\ 
 \hline
 $   \sigma(M_{\mu^{+}\mu^{-}})$  &                  6951.9 &                  4721.1 &                  89473.6 &                  48226.9 &                   305135 &                  6026.6 &                   453583 &                 0.66 &                   10.24\\ 
 \hline
 $                    MVA_{2f} $  &                   4430.7 &                   4075.1 &                    18137 &                  19007.4 &                   9303.8 &                   2059.2 &                  52582.5 &                      0.42 &                    18.56\\ 
 \hline
 $                    MVA_{4f} $ &                   2877.2 &                   2335.3 &                   3734.9 &                   5716.4 &                     3820 &                    201.8 &                  15808.4 &                     0.28 &                    21.05\\ 
 \hline
ISR photon veto &                    2875.3 &                   2334.2 &                     3733 &                   5716.4 &                      974 &                    201.8 &                  12959.4 &                     0.28 &                    22.85\\ 
 \hline\hline
 $\ms - 20$\,GeV$ < \mrec  < 450$\,GeV &                   2875.3 &                   2334.2 &                     3733 &                   5716.4 &                      974 &                    201.8 &                  12959.4 &                     0.28 &                    22.85\\ 
 \hline
 \end{tabular}
 \end{scriptsize}

\subcaption{ IDR-L, $\ms=100$ GeV}

 \begin{scriptsize}
 \begin{tabular}{|c| c| c| c| c| c| c| c| c| c|}
 \hline
   &   $             \ms=100 $  &  $              \mu^+\mu^- h $ &  $              4f_{l} $ &  $             4f_{sl} $ &  $              2f_{l} $ &  $            6f$ &  $B_{\mathrm{total}}$ &               $\epsilon$ &             $\Sigma$\\ 
\hline
no cut ($ \sigma\times \int\mathcal{L}dt$) &                  5915.8 &                  5519.2 &                1.69$\times 10^{7}$ &                  2.11$\times 10^{7}$ &                5.44$\times 10^{6}$&                 1.88$\times 10^{6}$ &                   9.06$\times 10^{7}$   &                        1 &        0.62          \\  
 \hline
$       M_{\mu^{+}\mu^{-}}^{FSR}  $ \& $   P^{T,FSR}_{\mu^{+}\mu^{-}} $  &                  4634.1 &                  4345.9 &                  64602.5 &                  51709.3 &                   509212 &                  1936.4 &                   631806 &                        0.78 &                  5.81\\ 
 \hline
 $   \sigma(M_{\mu^{+}\mu^{-}})$  &                   4485.7 &                   4237.3 &                  47988.4 &                  41353.1 &                   216868 &                  1779.5 &                   312226 &                 0.76 &                  7.97\\ 
 \hline
$                    MVA_{2f} $  &                   3806.3 &                   3628.5 &                   9741.5 &                  14999.4 &                   5837.9 &                    605.5 &                  34812.8 &                     0.64 &                    19.37\\ 
 \hline
$                    MVA_{4f} $  &                   3134.1 &                   2882.8 &                   4852.1 &                   7955.6 &                   3578.2 &                    281.6 &                  19550.2 &                     0.53 &                    20.81\\ 
 \hline
 ISR photon veto &                       3134.1 &                   2882.8 &                   4851.8 &                   7955.6 &                   1380.1 &                    281.6 &                  17351.9 &                     0.53 &                     21.9\\ 
 \hline\hline
 $\ms - 20$\,GeV$ < \mrec  < 450$\,GeV &                   3094.2 &                   2877.2 &                   4648.7 &                   7693.9 &                    654.1 &                    281.6 &                  16155.5 &                     0.52 &                     22.3\\ 
 \hline
 \end{tabular}
 \end{scriptsize}

\subcaption{ IDR-S, $\ms=100$ GeV}
 
 \begin{scriptsize}
 \begin{tabular}{|c| c| c| c| c| c| c| c| c| c|}
 \hline
   &   $             \ms=100 $  &  $              \mu^+\mu^- h $ &  $              4f_{l} $ &  $             4f_{sl} $ &  $              2f_{l} $ &  $            6f$ &  $B_{\mathrm{total}}$ &               $\epsilon$ &             $\Sigma$\\ 
 \hline
no cut ($ \sigma\times \int\mathcal{L}dt$) &                   5915.8 &                  5519.2 &                1.69$\times 10^{7}$ &                  2.11$\times 10^{7}$ &                5.44$\times 10^{6}$&                 1.88$\times 10^{6}$ &                   9.06$\times 10^{7}$   &                        1 &        0.62          \\  
 \hline
 $       M_{\mu^{+}\mu^{-}}^{FSR}  $ \& $   P^{T,FSR}_{\mu^{+}\mu^{-}} $   &                  4670.3 &                  4399.4 &                  64239.4 &                  52079.5 &                   507098 &                  2058.5 &                   629875 &                        0.79 &                  5.86\\ 
 \hline
 $    \sigma(M_{\mu^{+}\mu^{-}})$ &                  4548.8 &                  4316.0 &                  50915.9 &                  43444.9 &                   271645 &                   1935.1 &                   372257 &                 0.77 &                  7.41\\ 
 \hline
 $                       MVA_{2f} $ &                     3872 &                   3707.4 &                  10589.8 &                  15193.9 &                     6780 &                    716.7 &                  36987.8 &                     0.65 &                    19.16\\ 
 \hline
 $                     MVA_{4f} $  &                   3146.4 &                   2880.8 &                   5175.9 &                   7952.5 &                     4267 &                    318.6 &                  20594.6 &                     0.53 &                    20.42\\ 
 \hline
  ISR photon veto &                    3145.5 &                   2879.7 &                   5171.7 &                   7952.5 &                   1279.2 &                    318.5 &                  17601.6 &                     0.53 &                    21.84\\ 
 \hline\hline
 $\ms - 20$\,GeV$ < \mrec  < 450$\,GeV &                   3108.6 &                     2870 &                   4913.8 &                   7631.6 &                    496.5 &                    318.4 &                  16230.2 &                     0.53 &                    22.35\\ 
 \hline
 \end{tabular}
 \end{scriptsize}

\subcaption{  IDR-L, $\ms=300$ GeV}

 \begin{scriptsize}
 \begin{tabular}{|c| c| c| c| c| c| c| c| c| c|}
 \hline
   &   $             \ms=300 $ &  $              \mu^+\mu^- h $ &  $              4f_{l} $ &  $             4f_{sl} $ &  $              2f_{l} $ &  $            6f$ &  $B_{\mathrm{total}}$ &               $\epsilon$ &             $\Sigma$\\ 
\hline
no cut ($ \sigma\times \int\mathcal{L}dt$) &                 1558.8 &                  5519.2 &                1.69$\times 10^{7}$ &                  2.11$\times 10^{7}$ &                5.44$\times 10^{6}$&                 1.88$\times 10^{6}$ &                   9.06$\times 10^{7}$   &                        1 &        0.16         \\     
 \hline
$       M_{\mu^{+}\mu^{-}}^{FSR}  $ \& $   P^{T,FSR}_{\mu^{+}\mu^{-}} $ &                  1192.1 &                   1469.4 &                  65459.8 &                  31938.8 &                   451439 &                  3225.3 &                   553532 &                        0.76 &                  1.60\\ 
 \hline
 $    \sigma(M_{\mu^{+}\mu^{-}})$ &                  1178.2 &                  1361.9 &                  48910.5 &                  21582.5 &                   159095 &                  3070.4 &                   234020 &                 0.76&                  2.43\\ 
 \hline
  $                   MVA_{2f} $ &                    959.5 &                    475.6 &                   8182.1 &                   2854.9 &                    649.4 &                   1450.4 &                  13612.5 &                      0.62 &                     7.95\\ 
 \hline
 $                   MVA_{4f} $  &                    883.4 &                    308.5 &                   5318.2 &                   1798.9 &                    400.2 &                      993 &                   8818.9 &                     0.57 &                     8.97\\ 
 \hline
 ISR photon veto  &                     883.4 &                    304.1 &                   5308.8 &                   1787.5 &                    371.1 &                    992.9 &                   8764.4 &                     0.57 &                     8.99\\ 
 \hline\hline
 $\ms - 20$\,GeV$ < \mrec  < 450$\,GeV &                    877.7 &                    186.3 &                   4158.3 &                   1105.4 &                    128.7 &                    796.8 &                   6375.5 &                     0.56 &                    10.31\\ 
 \hline
 \end{tabular}
 \end{scriptsize}

\subcaption{  IDR-S, $\ms=300$ GeV}

 \begin{scriptsize}
 \begin{tabular}{|c| c| c| c| c| c| c| c| c| c|}
 \hline
   &   $             \ms=300 $  &  $              \mu^+\mu^- h $ &  $              4f_{l} $ &  $             4f_{sl} $ &  $              2f_{l} $ &  $            6f$ &  $B_{\mathrm{total}}$ &               $\epsilon$ &             $\Sigma$\\ 
 \hline
no cut ($ \sigma\times \int\mathcal{L}dt$) &                  1558.8 &                  5519.2 &                1.69$\times 10^{7}$ &                  2.11$\times 10^{7}$ &                5.44$\times 10^{6}$&                 1.88$\times 10^{6}$ &                   9.06$\times 10^{7}$   &                        1 &        0.16         \\  
 \hline
$       M_{\mu^{+}\mu^{-}}^{FSR}  $ \& $   P^{T,FSR}_{\mu^{+}\mu^{-}} $  &                  1206.7 &                  2012.1 &                  64036.7 &                  31028.2 &                   445395 &                  3410.5 &                   545882 &                        0.77 &                  1.63\\
\hline
 $    \sigma(M_{\mu^{+}\mu^{-}})  $ &                  1193.3 &                  1929.8 &                  50784.8 &                  22393.6 &                   205009 &                  3288.5 &                   283406 &                 0.77&                  2.24\\ 
 \hline
 $                    MVA_{2f} $&                    968.7 &                    498.2 &                   9220.5 &                   3008.8 &                   1470.4 &                   1597.7 &                  15795.5 &                      0.62 &                     7.48\\ 
 \hline
 $                    MVA_{4f} $ &                    873.3 &                      343 &                   5824.1 &                   1999.1 &                    795.1 &                     1064 &                  10025.2 &                     0.56 &                     8.37\\ 
 \hline
ISR photon veto &                   872.7 &                    339.7 &                   5809.9 &                   1991.3 &                    795.1 &                   1063.8 &                   9999.7 &                     0.56 &                     8.37\\ 
 \hline\hline
 $\ms - 20$\,GeV$ < \mrec  < 450$\,GeV &                      868 &                    198.5 &                   4429.9 &                     1123 &                    312.2 &                    853.7 &                   6917.3 &                     0.56 &                     9.84\\ 
 \hline
 \end{tabular}
 \end{scriptsize}
 \end{center} 
 \vspace{-0.5cm}
  \caption{Cut flow for $\ms = 20, 100$ and $300$\,GeV in IDR-L and IDR-S for $1600$\,fb$^{-1}$ of data with 
   $P(e^-,e^+)=(-80\%,+30\%)$.  The signal is normalised to
    $\sin^2{\theta}=1$. 
%The last row ($\mrec$) gives for information the event counts in the range $\ms - 20$\,GeV$ < \mrec < 450$\,GeV.
    }\label{table_cutflow}
 \end{table}

\begin{figure}[htb]
\begin{center}
\begin{subfigure}{0.45\textwidth}
  \includegraphics[width=\textwidth]{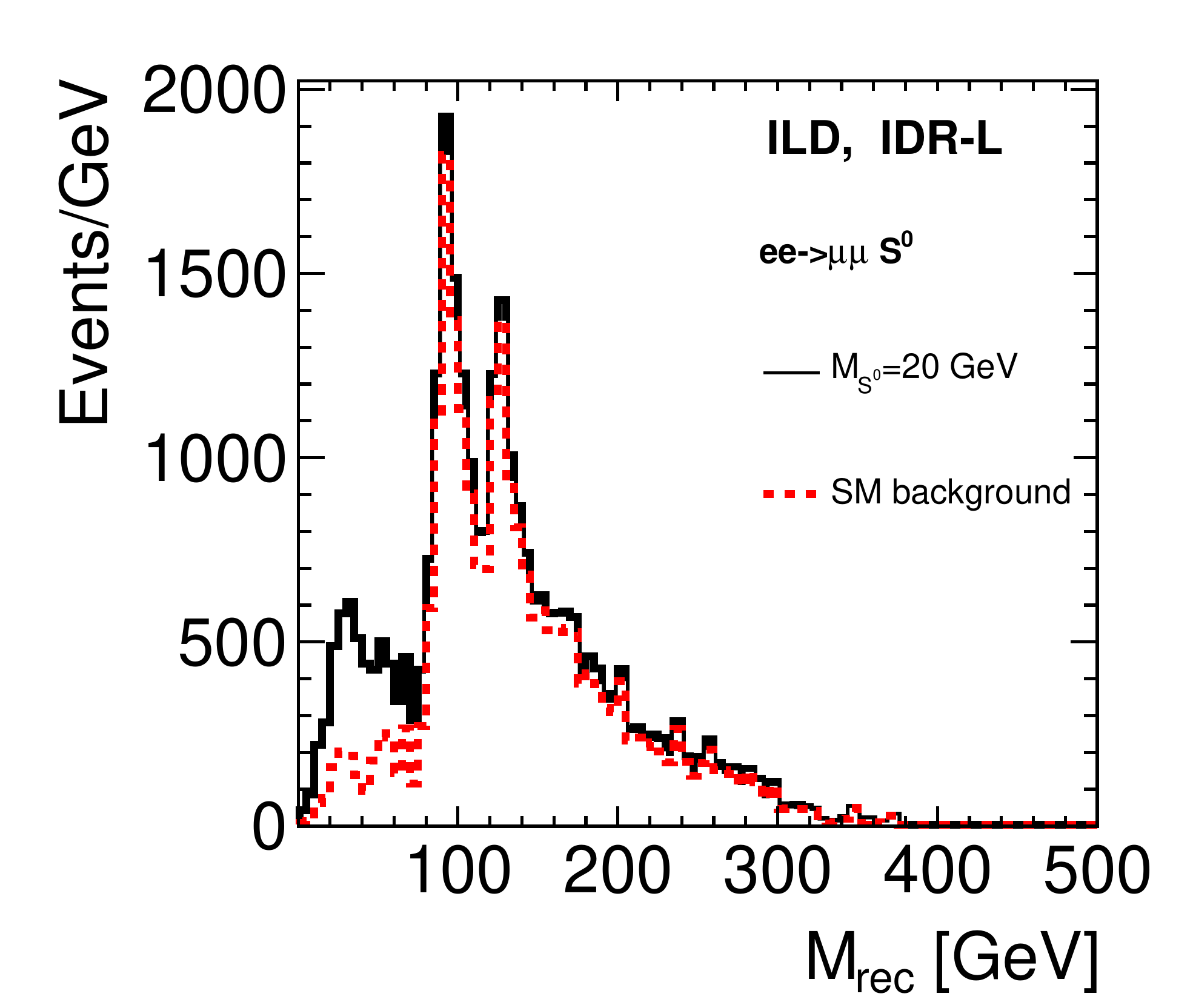}	
  \subcaption{}
\end{subfigure}
\begin{subfigure}{0.45\textwidth}
  \includegraphics[width=\textwidth]{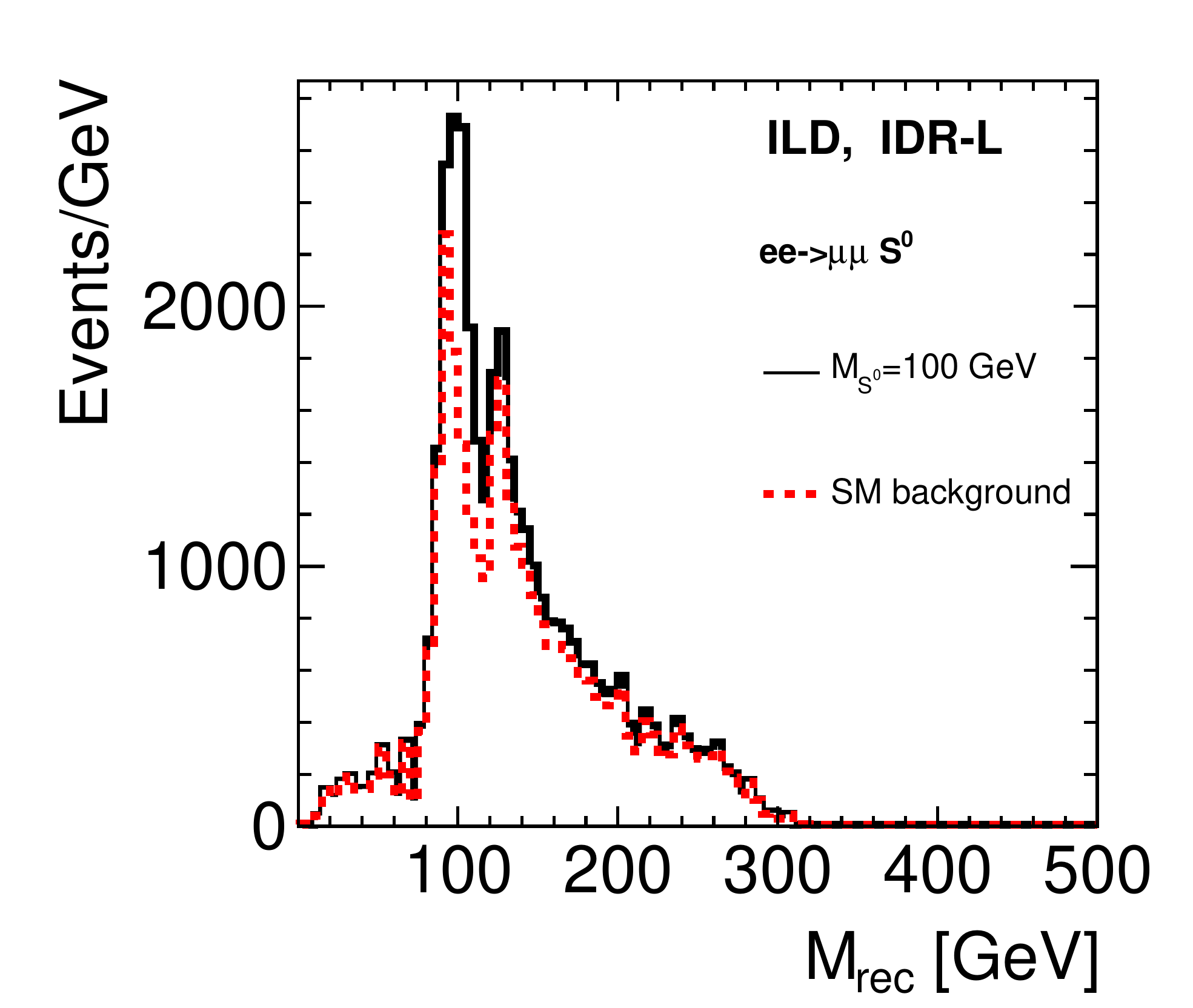}	
  \subcaption{}
\end{subfigure}
\hspace{0.03cm}
\begin{subfigure}{0.45\textwidth}
  \includegraphics[width=\textwidth]{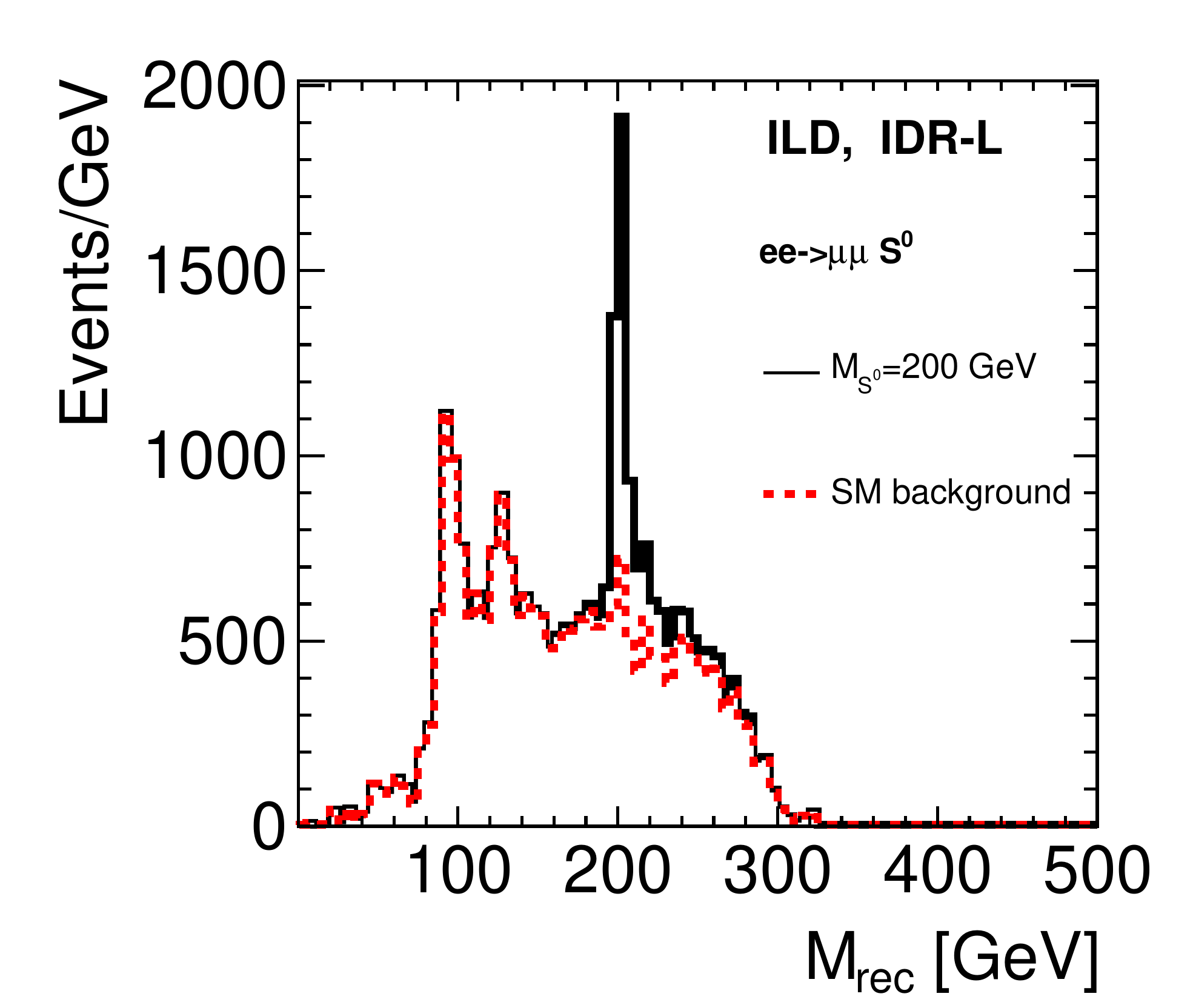}	
  \subcaption{}
\end{subfigure}
\begin{subfigure}{0.45\textwidth}
  \includegraphics[width=\textwidth]{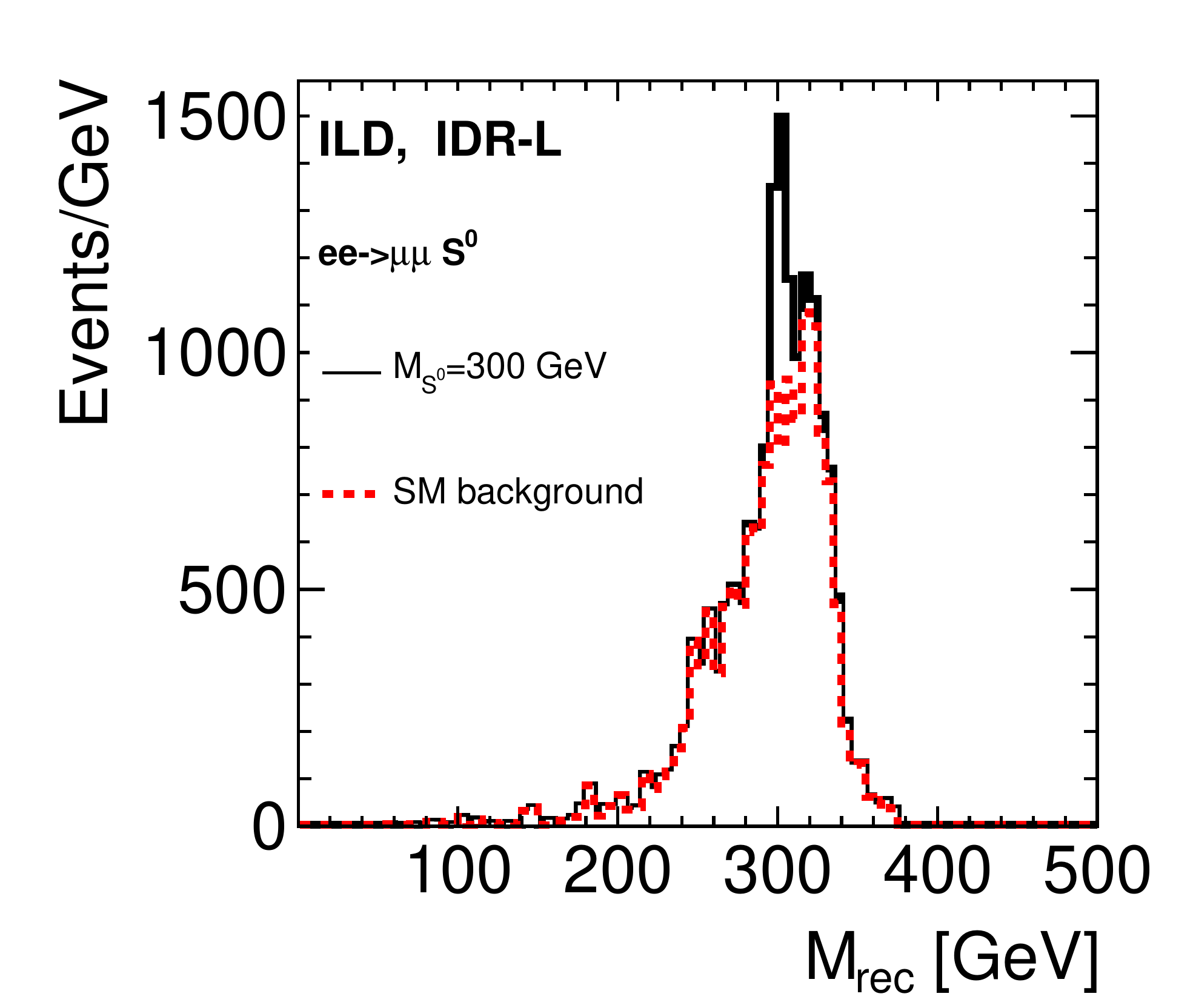}	
  \subcaption{}
\end{subfigure}
\end{center}
\caption{The recoil mass distributions after the ISR photon veto cut for signal and backgrounds 
within IDR-L model on the running scenario, when $\ms=20, 100, 200, 300$ GeV and $\sin^2\theta=1$.}
\label{fig_recoil}
\end{figure}

The recoil mass distributions after the full selection, which will be the input to the final sensitivity 
calculation, are shown in Fig.~\ref{fig_recoil} for the examples 
of $\ms = 20, 100, 200$ and $300$\,GeV. In all cases, the signal distributions, which are normalised to 
$\sin^2{\theta}=1$, peak around the corresponding 
value of $\ms$. The smaller $\ms$, the wider the recoil peak becomes due to the increased probability of ISR.
The different shapes of the background distributions originate from the $\ms$-dependent optimisation of the 
MVA-based part of the selection. For all low and medium $\ms$, the peaks
of the $Z$ and the 125 GeV Higgs bosons can be clearly seen in the background distribution.   

\begin{figure}[htb]
\begin{center}
\begin{subfigure}{0.45\textwidth}
  \includegraphics[width=\textwidth]{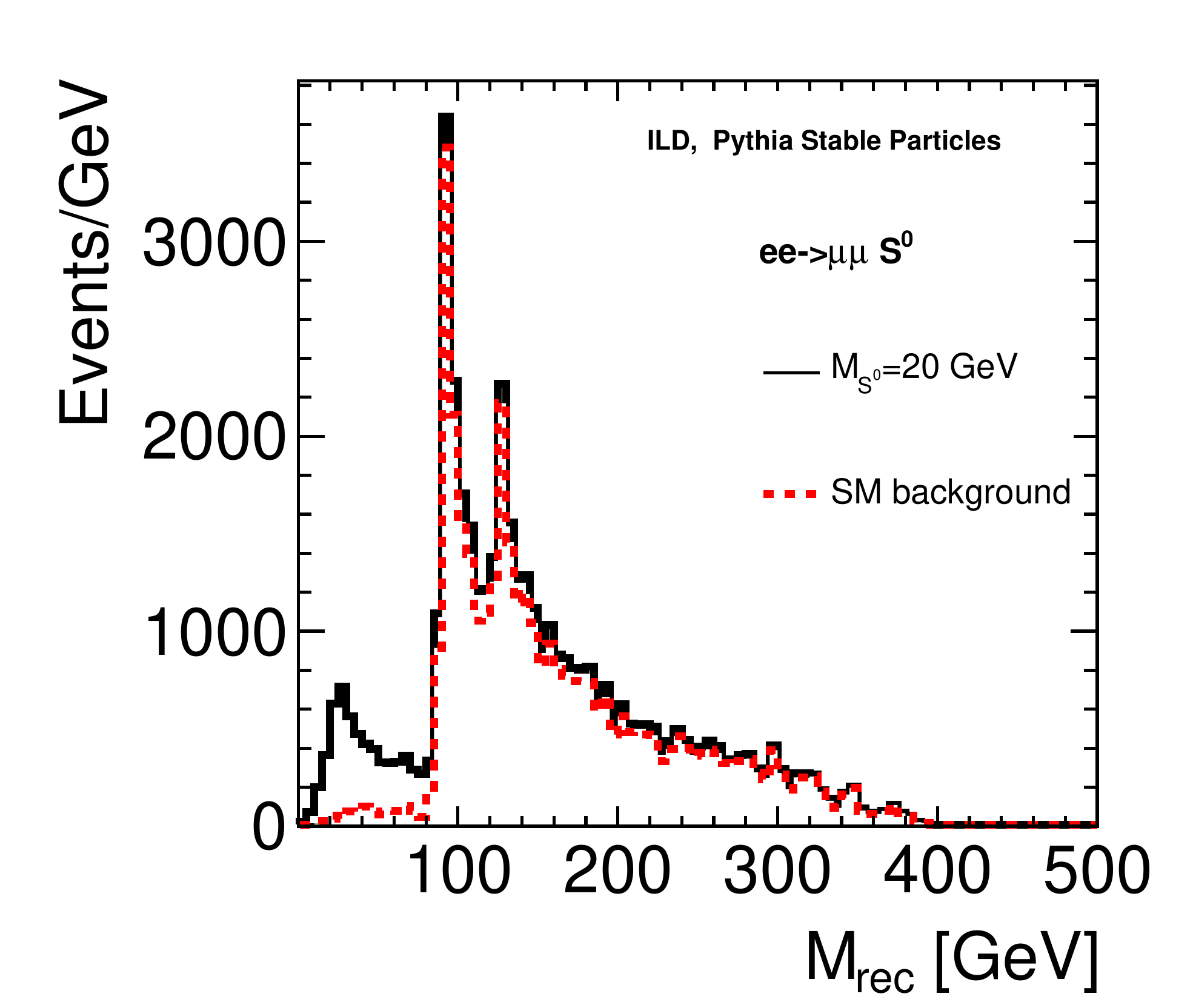}	
  \subcaption{}
\end{subfigure}
\hspace{0.03cm}
\begin{subfigure}{0.45\textwidth}
  \includegraphics[width=\textwidth]{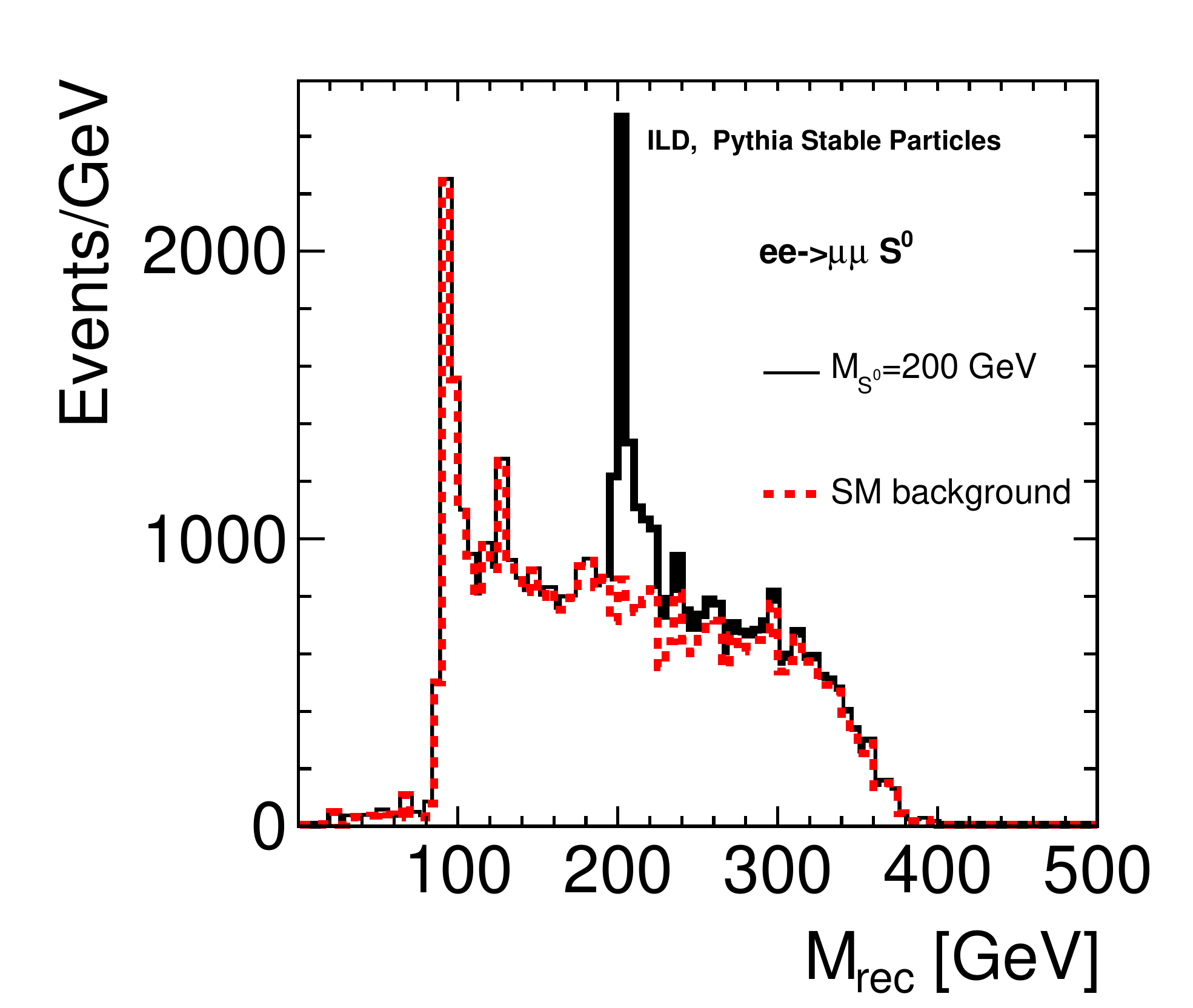}	
  \subcaption{}
\end{subfigure}
\end{center}
 \caption{The recoil mass distributions after the ISR photon veto cut for signal and backgrounds 
within Pythia stable particles, when $\ms=20, 200$ GeV and $\sin^2\theta=1$.}
\label{fig_recoil_py}
\end{figure}

Figure~\ref{fig_recoil_py} shows two examples of the recoil mass distributions obtained from running the 
whole analysis on the ``Pythia-stable-particle'' level, i.e.\ at truth level after the hadronisation step, for 
$\ms=20$ and $200$\,GeV and again $\sin^2{\theta}=1$. In the case of $\ms=200$\,GeV, the recoil distributions
at ``Pythia-stable-particle'' level and after full reconstruction are rather similar. The peaks are sharper 
at truth level and the overall events numbers are higher, but in particular in the region of the signal peak,
the signal-to-background ratio is very similar between truth and reconstruction levels. Thus we expect also
the final sensitivity to be similar. The situation is quite different for the low mass case of $\ms=20$\,GeV:
In the important recoil mass range below $\simeq 90$\,GeV, the signal-to-background ratio is much worse after 
full reconstruction compared to the truth level.

In order to shed further light on this fact, Table~\ref{table_90_region} shows the background decompostion and 
the signal significance
for $\ms=20$\,GeV for the whole recoil mass range and for the region below $90$\,GeV, both at ``Pythia-stable-particle'' 
and at full reconstruction level (IDR-L). The signal significance in the relevant mass window drops from $44$ on 
truth level to $32$ on reconstruction level
which is partially due to $25\%$ less signal events being selected, and partially due to the background increasing 
by more than a factor $2$. The largest increase in background originates from the $e^{+}e^{-} \to \mu^{+}\mu^{-} \gamma$ 
process ($2f_{l}$ category), followed by the $ZZ\to \mu^{+}\mu^{-} q \bar{q}$ ($4f_{sl}$ category). Both are harder 
to reject due to imperfections in the identification of the ISR photon, due missing ISR real photons and due to 
misidentifying photons from jets as ISR, which leads to a migration into the low recoil mass region.
Thus an important conclusion from this study is that the most important detector and reconstruction performance 
aspect for this channel is not the muon identification or momentum resolution, but the correct identification 
of ISR photons.

\begin{table}[htb]
 \begin{center}
 \begin{small}
 \begin{tabular}{|c|c|c|c|c|c|c|c|c|}
\hline
Pythia case &   $  \ms=20 $  &  $  \mu^+\mu^- h $ &  $  4f_{l} $ &  $  4f_{sl} $ & $  2f_{l} $ & $6f_{l}$ & total $B$ &    $\Sigma$ \\ 
\hline
$\mrec$ = [0-450] GeV  &    4216.0    &      3256.9 &      6809.9 &          9910 &     31.8 &   533.3  &   20541.9 &           26.79\\ 
\hline
$\mrec$ = [0-90] GeV  &       2667.2 &            0 &         491.6 &         462.1 &       0 &      0 &   953.7 &            44.32\\              
\hline

\hline
IDR-L  &   $  \ms=20 $  &  $  \mu^+\mu^- h $ &  $  4f_{l} $ &  $  4f_{sl} $ & $  2f_{l} $  &  $6f_{l}$ & total $B$  &   $\Sigma$ \\ 
\hline
$\mrec$ = [0-450] GeV  &        2906.7 &       2347.2 &       3601.6 &        5674.5 &      1058.4 &     177.9 &  12859.7 &                23.15\\ 
\hline
$\mrec$ = [0-90] GeV &        2034.5 &          7.8 &        486.1 &         811.4 &       639.9 &      0 &  1945.2 &              32.25\\   
\hline
 \end{tabular}
 \end{small}
 \end{center}
  \caption{Background decompostion for $\ms=20$\,GeV, both at ``Pythia-stable-particle'' 
and at full reconstruction level (IDR-L).}\label{table_90_region}
 \end{table}

\section{Results}
The recoil mass distributions as shown in Fig.~\ref{fig_recoil} are input to the final sensitivity calculation 
via a frequentist approach based on fractional event counting. For each considered mass of
of the scalar, the value of $\sin^2{\theta}$ which is expected to be excluded at the $95\%$ confidence level ($CL$) 
 in absence of a signal is calculated using the $CL_S$-method~\cite{Read:2002hq}. The calculation is performed
using the package \textsc{TSysLimit}, which was originally written for leptoquark searches at 
\textsc{HERA}~\cite{Aktas:2005pr}.
 
\begin{figure}[htb]
\begin{center}

\begin{subfigure}{0.475\textwidth}
  \includegraphics[width=\textwidth]{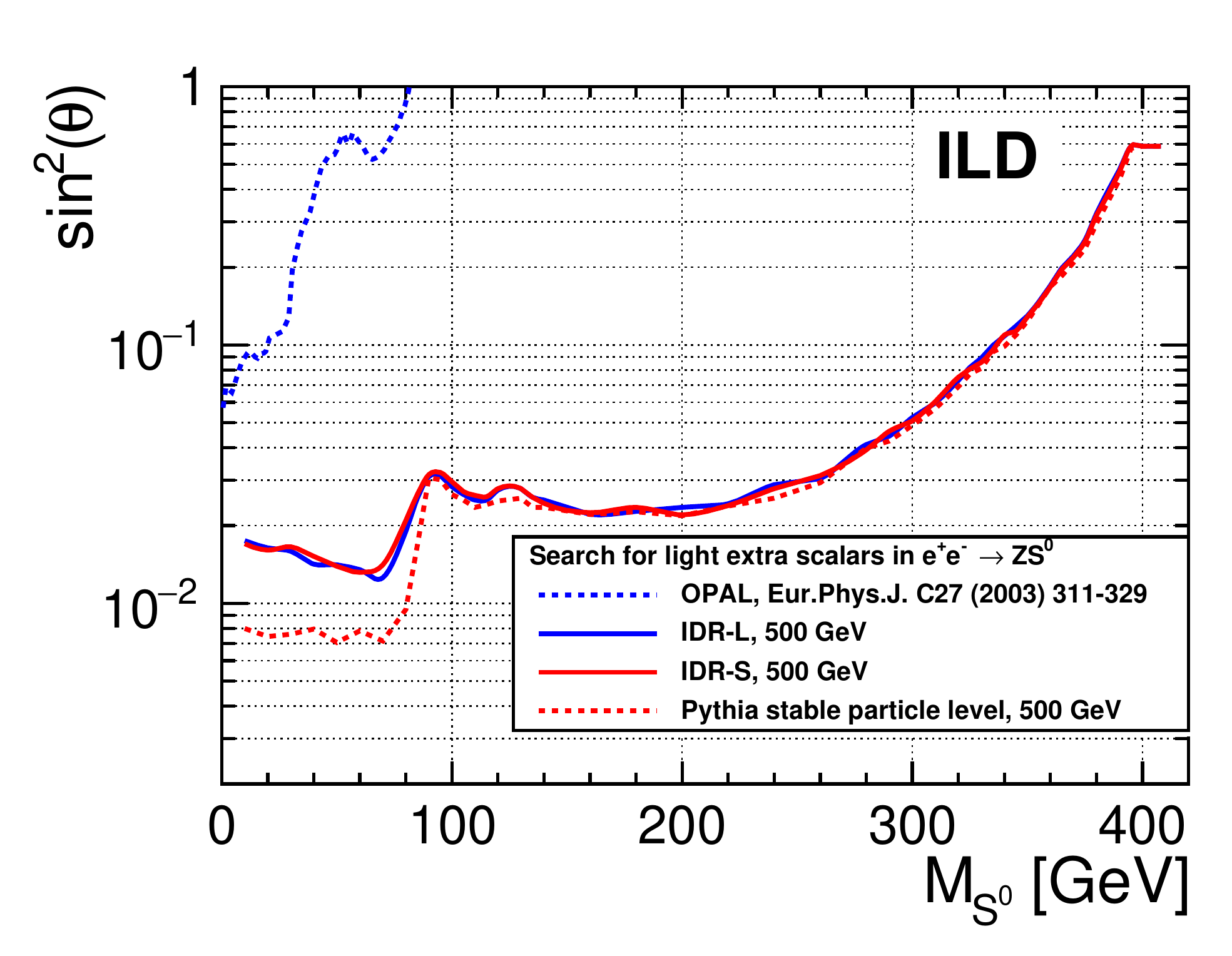} 
  \subcaption{\label{fig_exclusion_limits:500}}
\end{subfigure}
\hspace{0.03cm}
\begin{subfigure}{0.475\textwidth}
  \includegraphics[width=\textwidth]{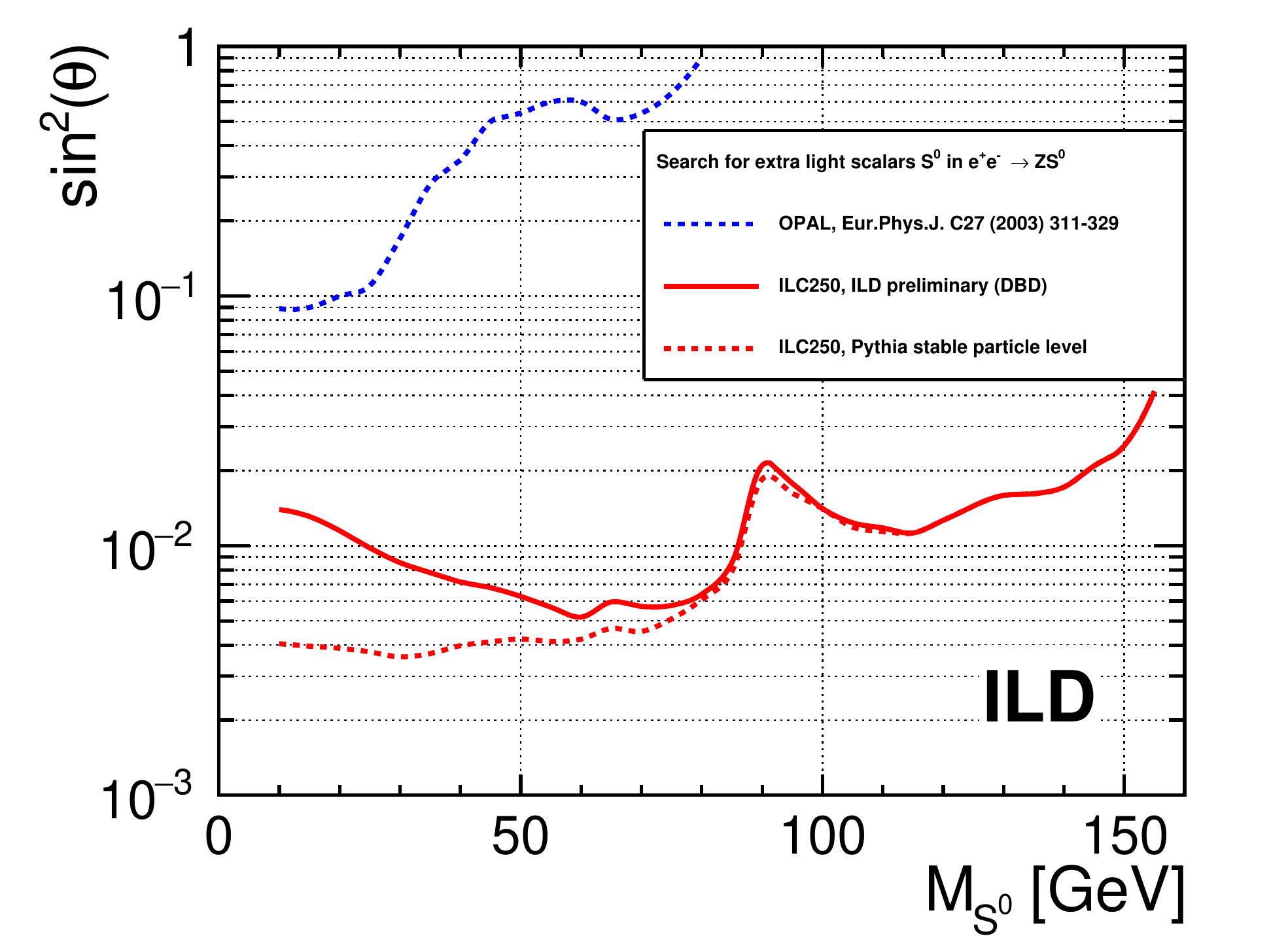} 
  \subcaption{\label{fig_exclusion_limits:250}}
\end{subfigure}
\end{center}
  \caption{(a) Expected sensitivity of ILC500 at the $95\%$ $CL$ for IDR-L and IDR-S compared to 
  the existing limit from LEP. (b) Corresponding result for the ILC run at $\sqrt{s}$ = 250 GeV.
  In both cases, the ``Pythia-stable-particle'' level result is also shown.
  }
  \label{fig_exclusion_limits}
\end{figure}

Figure~\ref{fig_exclusion_limits:500} shows the resulting $95\%$ $CL$ sensitivity on $\sin^2{\theta}$ as 
a function of $\ms$ for ILC operation at $\sqrt{s}=500$ GeV. Also shown, as blue dashed line, is the corresponding limit 
observed by the OPAL collaboration~\cite{Abbiendi}.
The OPAL result is based on a combination of the $Z \to \ee$ and $Z \to \mm$ channels while the analysis 
in this note only uses the $Z \to \mm$ events. In order to compare directly, the ILD projections have been
scaled by a factor $1/\sqrt{2}$, assuming that the sensitivity for the $Z \to \ee$  channel is similar 
to the $Z \to \mm$ channel 
analysed here, which is supported by the roughly similar performance of the two channels in the 
measurement of the total $ZH(125)$ cross section via the same recoil technique~\cite{Yan:2016xyx}.
The ILC would cover large additional parameter space, both in $\sin^2{\theta}$ as well as in $\ms$, compared 
to the OPAL result. 
There is no significant difference between IDR-L (the blue solid line) and IDR-S (the red solid line). 
The red dashed line shows the result obtained on the ``Pythia-stable-particle'' level, i.e.\ from the recoil
distributions shown in Fig.~\ref{fig_recoil_py}. At low masses, a clear difference to the full reconstruction
result can be seen, as expected based on the discussion of Table~\ref{table_90_region}. For masses above the
$Z$ resonance, the full reconstruction result is as good as the truth level expectation.

Although the $500$\,GeV projection promisses a significant improvement over the OPAL result, 
even stronger constraints on $\sin^2{\theta}$ can be obtained for the kinematically accessible
masses at the inital ILC run at $250$\,GeV. Figure~\ref{fig_exclusion_limits:250} the corresponding (preliminary)
projection~\cite{Wang:2019mzd} for $\sqrt{s} = 250$\,GeV with the canonic ILC running assumptions for this energy (namely $2$\,ab$^{-1}$ with  $|P(e^{+}e^{-})| = (30\%, 80\%)$ shared according to
$f(-+, +-, ++, --) = (45\%, 45\%, 5\%, 5\%)$), again compared to the OPAL result and the
``Pythia-stable-particle'' level. Also in this case, the impact of detector and reconstruction performance 
is largest for scalar masses below the $Z$ pole, due to the imperfections in the ISR reconstruction discussed above.

\section{Conclusions}
The potential of the ILC to search for new scalars in a model-independent way via the recoil technique
has been investigated in full simulation of the ILD concept. 
Two different variants of the ILD detector have been compared at $\sqrt{s} = 500$ GeV, assuming  
an integrated luminosity of $4000$\,fb$^{-1}$ and beam polarisations of $|P(e^{+}e^{-})| = (30\%, 80\%)$, 
shared between the $(-+, +-, ++, --)$ configurations as $(40\%, 40\%, 10\%, 10\%)$, 
 i.e.\ $(1600, 1600, 400, 400)$\,fb$^{-1}$. Expected sensitivities at 95\% CL for the 
cross section scale factor with respect to\ the SM Higgs, $\sin^2{\theta}$, are shown for scalar masses between
$10$ and $410$\,GeV. 
They are one or two orders of magnitudes more sensitive than those obtained at LEP, and covering 
substantial new phase space. No obvious difference between IDR-L and IDR-S is observed for this analysis. 
The most significant limitation comes from the ISR photon identification.

\section*{Acknowledgements}
We would like to thank the LCC generator working group and the ILD software
working group for providing the simulation and reconstruction tools and producing
the Monte Carlo samples used in this study. This work has benefited from
computing services provided by the ILC Virtual Organization, supported by the
national resource providers of the EGI Federation and the Open Science GRID. We
are grateful for the support from Collaborative Research Center SFB676 of the
Deutsche Forschungsgemeinschaft (DFG), Particles, Strings and the Early Universe,
project B1. Y.W. is supported by High-level Scientific Research Foundation for the introduction of talent in Inner Mongolia Normal University in China, and an International Postdoctoral Exchange
Fellowship Program between the Office of the National Administrative Committee
of Postdoctoral Researchers of China (ONACPR) and DESY.

\appendix
\section{MVA output cut values for all scalar masses}
In this appendix, the MVA cuts for different scalar masses are shown in Table \ref{table_cut_masses}. These cuts are used for both IDR-L and IDR-S.
\begin{table}[!h]
 \begin{center}
 \begin{small}
 \begin{tabular}{|c |c| c| c| c| c| c| c| c| c| c| }
 \hline
      $\ms$ & 10 & 20 & 30 & 40 & 50 & 60 & 70 & 80 & 85 & 90  \\
 \hline
      $MVA_{2f}$ & [0.8,1] & [0.85,1] & [0.85,1] & [0.9,1] & [0.85,1] & [0.85,1] & [0.9,1] & [0.9,1] & [0.9,1] & [0.9,1]  \\
  \hline
      $MVA_{4f}$ & [0.4,1] & [0.5,1] & [0.4,1] & [0.4,1] & [0.3,1] &[0.4,1] & [0.4,1] & [0.3,1] & [0.4,1] & [0.4,1]  \\
 \hline
      $\ms$ &  95 & 100 & 105 & 110 & 115 & 120 & 130 & 135 & 140 & 160  \\
 \hline
   $MVA_{2f}$ & [0.9,1] & [0.9,1] & [0.9,1] & [0.9,1] & [0.9,1] & [0.9,1] & [0.85,1] & [0.9,1] & [0.9,1] & [0.9,1]  \\
   \hline
      $MVA_{4f}$ & [0.4,1] & [0.4,1] & [0.4,1] & [0.4,1] & [0.4,1] &[0.4,1] & [0.4,1] & [0.4,1] & [0.4,1] & [0.4,1]  \\
 \hline
       $\ms$ & 180 & 200 & 220 & 240 & 260 & 280 & 290 & 300 & 310 & 320  \\
  \hline
   $MVA_{2f}$ & [0.9,1] & [0.9,1] & [0.9,1] & [0.9,1] & [0.9,1] & [0.75,1] & [0.9,1] & [0.95,1] & [0.95,1] & [0.95,1]  \\
   \hline
      $MVA_{4f}$ & [0.4,1] & [0.5,1] & [0.4,1] & [0.5,1] & [0.5,1] &[0.5,1] & [0.5,1] & [0.5,1] & [0.5,1] & [0.6,1]  \\
 \hline
       $\ms$ & 325 & 330 & 335 & 340 & 345 & 350 & 355 & 360 & 365 & 370  \\
  \hline
   $MVA_{2f}$ & [0.95,1] & [0.95,1] & [0.95,1] & [0.95,1] & [0.95,1] & [0.95,1] & [0.95,1] & [0.95,1] & [0.95,1] & [0.95,1]  \\
  \hline
      $MVA_{4f}$ & [0.6,1] & [0.6,1] & [0.6,1] & [0.6,1] & [0.6,1] &[0.6,1] & [0.6,1] & [0.6,1] & [0.4,1] & [0.4,1]  \\
 \hline
       $\ms$ & 375 & 380 & 385 & 390 & 395 & 400 & 405 & 408 & - &  - \\
  \hline
   $MVA_{2f}$ & [0.75,1] & [0.75,1] & [0.75,1] & [0.75,1] & [0.75,1] & [0.75,1] & [0.75,1] & [0.75,1] & - & - \\
   \hline
      $MVA_{4f}$ & [0.5,1] & [0.5,1] & [0.5,1] & [0.5,1] & [0.5,1] &[0.5,1] & [0.5,1] & [0.5,1] & - & -  \\
 \hline
 \end{tabular}
 \end{small}
 \end{center}

  \caption{The MVA cuts for different scalar masses. These cuts are used for both IDR-L and IDR-S.}\label{table_cut_masses}
 \end{table}


\begin{thebibliography}{99}
\bibitem{ATLAS}{The ATLAS and CMS collaboration, "Measurements of the Higgs boson production and decay rates
and constraints on its couplings from a combined ATLAS and CMS analysis of the LHC pp collision
data at $\sqrt{S} = 7$ and 8 TeV", JHEP 08 (2016) p.045.}

\bibitem{aggleton}{ R. Aggleton et al., "Review of LHC experimental results on low mass bosons in multi Higgs models",
JHEP 02 (2017) p. 035.}

  \bibitem{Barate:2003sz} 
  R.~Barate {\itshape et al.} [ALEPH and DELPHI and L3 and OPAL Collaborations and LEP Working Group for Higgs boson searches],
  ``Search for the standard model Higgs boson at LEP,''
  Phys.\ Lett.\ B {\bfseries 565}, 61 (2003), are provided
  %doi:10.1016/S0370-2693(03)00614-2
  [hep-ex/0306033].
  %%CITATION = doi:10.1016/S0370-2693(03)00614-2;%%
  %2712 citations counted in INSPIRE as of 21 Nov 2019


\bibitem{Schael:2006cr} 
  S.~Schael {\itshape et al.} [ALEPH and DELPHI and L3 and OPAL Collaborations and LEP Working Group for Higgs Boson Searches],
  ``Search for neutral MSSM Higgs bosons at LEP,''
  Eur.\ Phys.\ J.\ C {\bfseries 47}, 547 (2006)
  %doi:10.1140/epjc/s2006-02569-7
  [hep-ex/0602042].
  %%CITATION = doi:10.1140/epjc/s2006-02569-7;%%
  %918 citations counted in INSPIRE as of 21 Nov 2019
 
 \bibitem{Aad:2015wra} 
  G.~Aad {\itshape et al.} [ATLAS Collaboration],
  ``Search for a CP-odd Higgs boson decaying to Zh in pp collisions at $\sqrt{s} = 8$ TeV with the ATLAS detector,''
  Phys.\ Lett.\ B {\bfseries 744}, 163 (2015)
  %doi:10.1016/j.physletb.2015.03.054
  [arXiv:1502.04478 [hep-ex]].
  %%CITATION = doi:10.1016/j.physletb.2015.03.054;%%
  %134 citations counted in INSPIRE as of 21 Nov 2019

\bibitem{Khachatryan:2015lba} 
  V.~Khachatryan {\itshape et al.} [CMS Collaboration],
  ``Search for a pseudoscalar boson decaying into a Z boson and the 125 GeV Higgs boson in $\Plp\Plm b\overline{b}$ final states,''
  Phys.\ Lett.\ B {\bfseries 748}, 221 (2015)
  %doi:10.1016/j.physletb.2015.07.010
  [arXiv:1504.04710 [hep-ex]].
  %%CITATION = doi:10.1016/j.physletb.2015.07.010;%%
  %91 citations counted in INSPIRE as of 21 Nov 2019
  
  \bibitem{Khachatryan:2016are} 
  V.~Khachatryan {\itshape et al.} [CMS Collaboration],
  ``Search for neutral resonances decaying into a Z boson and a pair of b jets or $\tau$ leptons,''
  Phys.\ Lett.\ B {\bfseries  759}, 369 (2016)
  %doi:10.1016/j.physletb.2016.05.087
  [arXiv:1603.02991 [hep-ex]].
  %%CITATION = doi:10.1016/j.physletb.2016.05.087;%%
  %56 citations counted in INSPIRE as of 21 Nov 2019
  
\bibitem{Aaboud:2018eoy} 
  M.~Aaboud {\itshape et al.} [ATLAS Collaboration],
  ``Search for a heavy Higgs boson decaying into a $Z$ boson and another heavy Higgs boson in the $\ell\ell bb$ final state in $pp$ collisions at $\sqrt{s}=13$ TeV with the ATLAS detector,''
  Phys.\ Lett.\ B {\bfseries 783}, 392 (2018)
  %doi:10.1016/j.physletb.2018.07.006
  [arXiv:1804.01126 [hep-ex]].
  %%CITATION = doi:10.1016/j.physletb.2018.07.006;%%
  %31 citations counted in INSPIRE as of 21 Nov 2019

  \bibitem{Sirunyan:2019xls} 
  A.~M.~Sirunyan {\itshape et al.} [CMS Collaboration],
  ``Search for a heavy pseudoscalar boson decaying to a Z and a Higgs boson at $\sqrt{s} =$ 13 TeV,''
  Eur.\ Phys.\ J.\ C {\bfseries 79}, no. 7, 564 (2019)
  %doi:10.1140/epjc/s10052-019-7058-z
  [arXiv:1903.00941 [hep-ex]].
  %%CITATION = doi:10.1140/epjc/s10052-019-7058-z;%%
  %12 citations counted in INSPIRE as of 21 Nov 2019
  
\bibitem{Sirunyan:2019wrn} 
  A.~M.~Sirunyan {\itshape et al.} [CMS Collaboration],
  ``Search for new neutral Higgs bosons through the H$\to$ ZA $\to \ell^{+}\ell^{-} \mathrm{b\bar{b}}$ process in pp collisions at $\sqrt{s} =$ 13 TeV,''
  arXiv:1911.03781 [hep-ex].
  %%CITATION = ARXIV:1911.03781;%%
  
    
\bibitem{Abbiendi}{ G. Abbiendi , "Decay mode independent searches for new scalar bosons with the OPAL detector at
LEP", Eur. Phys. J. C27 (2003) p. 311-329.}

\bibitem{Asner}{ D. M. Asner et al., "ILC Higgs White Paper", Proceedings, CSS2013 (2013) Minneapolis, USA,
July 29-August 6, 2013.}

\bibitem{theo}{P. Drechsel, G. Moortgat-Pick, and G. Weiglein, "Sensitivity of the ILC to light Higgs masses", 	arXiv:1801.09662[hep-ph].}

\bibitem{Wang:2018fcw}{Y.~Wang, J. List,and M. Berggren,
  "Search for Light Scalars Produced in Association with
                        Muon Pairs for $\sqrt{s}$ = 250 GeV at the ILC", 
  arXiv:1801.08164 [hep-ex].}
  
\bibitem{Wang:2019mzd}{Y.~Wang, J. List,and M. Berggren,
  "Search for Extra Scalars Produced in Association with
                        Muon Pairs at the ILC", 
  arXiv:1902.06118 [hep-ex].}
  
\bibitem{Wang:2018ichep}{Y.~Wang, J. List,and M. Berggren,
  "Search for Light Scalars Produced in Association
with a $Z$ boson at the 250 GeV stage of the ILC",
  ICHEP2018 Proceeding, PoS(ICHEP2018) 630.}
  
\bibitem{Thomson:2015jda}
  M.~Thomson,
  %``Model-independent measurement of the e$^{{+}}$ e$^{-}$ $\rightarrow $ HZ cross section at a future e$^{{+}}$ e$^{-}$ linear collider using hadronic Z decays,''
  Eur.\ Phys.\ J.\ C {\bfseries 76} (2016) no.2,  72
  doi:10.1140/epjc/s10052-016-3911-5
  [arXiv:1509.02853 [hep-ex]].
  %%CITATION = doi:10.1140/epjc/s10052-016-3911-5;%%
  %31 citations counted in INSPIRE as of 14 Apr 2020
  
  \bibitem{Yan:2016xyx}
  J.~Yan, S.~Watanuki, K.~Fujii, A.~Ishikawa, D.~Jeans, J.~Strube, J.~Tian and H.~Yamamoto,
  %``Measurement of the Higgs boson mass and $e^+e^- \to ZH$ cross section using $Z \to \mu^+\mu^-$ and $Z \to e^+ e^-$ at the ILC,''
  Phys.\ Rev.\ D {\bfseries 94} (2016) no.11,  113002
  doi:10.1103/PhysRevD.94.113002
  [arXiv:1604.07524 [hep-ex]].
  %%CITATION = doi:10.1103/PhysRevD.94.113002;%%
  %26 citations counted in INSPIRE as of 14 Apr 2020

  
  \bibitem{ILD:2020qve}
  H.~Abramowicz {\itshape et al.} [ILD Concept Group],
  ``International Large Detector: Interim Design Report,''
  arXiv:2003.01116 [physics.ins-det].
  
  


\bibitem{Behnke:2013lya}
  T.~Behnke {\itshape et al.},
  ``The International Linear Collider Technical Design Report - Volume 4: Detectors,''
  arXiv:1306.6329 [physics.ins-det].
  
\bibitem{Kilian:2007gr} 
  W.~Kilian, T.~Ohl and J.~Reuter,
  "WHIZARD: Simulating Multi-Particle Processes at LHC and ILC",
  Eur.\ Phys.\ J.\ C {\bfseries 71}, 1742 (2011).
  
  \bibitem{Sjostrand:2006za}
  T.~Sjostrand, S.~Mrenna and P.~Z.~Skands,
  %``PYTHIA 6.4 Physics and Manual,''
  JHEP {\bfseries  0605} (2006) 026
  doi:10.1088/1126-6708/2006/05/026
  [hep-ph/0603175].
  
\bibitem{ilcsoft:2019}
``ILCSOFT home page,''  \url{http://ilcsoft.desy.de/
portal}.




  
\bibitem{Thomson:2009rp}
M.~Thomson,
``Particle Flow Calorimetry and the PandoraPFA Algorithm,''
Nucl. Instrum. Meth. A \textbf{611} (2009), 25-40
doi:10.1016/j.nima.2009.09.009
[arXiv:0907.3577 [physics.ins-det]].
%355 citations counted in INSPIRE as of 16 Apr 2020
  
  %\cite{Wendt:2007iw}
\bibitem{Wendt:2007iw}
O.~Wendt, F.~Gaede and T.~Kr\"amer,
``Event Reconstruction with MarlinReco at the ILC,''
Pramana \textbf{69} (2007), 1109-1114
doi:10.1007/s12043-007-0237-8
[arXiv:physics/0702171 [physics]].
%28 citations counted in INSPIRE as of 14 Apr 2020
  
  
  \bibitem{junping:ilt}
	Junping Tian, Claude D\"urig, isolated lepton finder, 2015, \url{https://github.com/iLCSoft/MarlinReco/tree/master/Analysis/IsolatedLeptonTagging}.


\bibitem{MVA}{TMVA home page https://root.cern/tmva.}

%\cite{Brun:1997pa}
\bibitem{Brun:1997pa}
R.~Brun and F.~Rademakers,
``ROOT: An object oriented data analysis framework,''
Nucl.\ Instrum.\ Meth.\ A \textbf{389} (1997), 81-86
doi:10.1016/S0168-9002(97)00048-X
%2277 citations counted in INSPIRE as of 31 Mar 2020





  
\bibitem{Read:2002hq}
  A.~L.~Read,
  ``Presentation of search results: The CL(s) technique,''
  J.\ Phys.\ G {\bfseries 28} (2002) 2693.
  doi:10.1088/0954-3899/28/10/313  
  
%\cite{Aktas:2005pr}
\bibitem{Aktas:2005pr}
A.~Aktas \textit{et al.} [H1],
``Search for leptoquark bosons in ep collisions at HERA,''
Phys.\ Lett.\ B \textbf{629} (2005), 9-19
doi:10.1016/j.physletb.2005.09.048
[arXiv:hep-ex/0506044 [hep-ex]].
%70 citations counted in INSPIRE as of 01 Apr 2020  

  
\end{thebibliography}
\end{document}